\documentclass[useAMs,a4paper, twocolumn]{mn2e}
\usepackage{savesym}
\usepackage{graphicx}
\expandafter\let\csname equation*\endcsname\relax
  \expandafter\let\csname endequation*\endcsname\relax 
\usepackage{subfig}
\usepackage{amsmath}
\usepackage{amssymb}
\usepackage{verbatim}
\usepackage[yyyymmdd,hhmmss]{datetime}
\usepackage{array}
\usepackage{times}
\usepackage[total={17.8cm,24.0cm},centering]{geometry} 
\usepackage{color}

\newcommand{\beq}{\begin{equation}}
\newcommand{\eeq}{\end{equation}}

\title[Analytical intra-ISCO accretion solutions]{ Accretion within the innermost stable circular orbit: analytical thermodynamic  solutions in the adiabatic limit }
\author [Andrew Mummery, Steven Balbus]{Andrew Mummery\thanks{E-mail:
andrew.mummery@physics.ox.ac.uk}, Steven Balbus
\\
Oxford Theoretical Physics, Beecroft Building,  Clarendon Laboratory, Parks Road, Oxford, OX1 3PU, United Kingdom}
\begin{document}

\date{}

\pagerange{\pageref{firstpage}--\pageref{lastpage}} \pubyear{2022}

\maketitle

\label{firstpage}

\begin{abstract} 
We present  analytical solutions for the thermodynamic (temperature, pressure, density, etc.)  properties of thin accretion flows in the region within the innermost stable circular orbit (ISCO) of a Kerr black hole, the first analytical solutions of their kind. These solutions are constructed in the adiabatic limit and neglect radiative losses, an {idealisation} valid for a restricted region of parameter space. We highlight a number of remarkable properties of these solutions, including that these solutions cool for radii $r_I/2 < r < r_I$, before increasing in temperature for $0 < r < r_I/2$, independent of black hole spin and assumptions regarding the equation of state of the accretion flow. The radiative temperature of these solutions can, for some values of the free parameters of the theory,   peak within the ISCO and not in the main body of the disc.  These solutions represent a fundamentally new class of analytical accretion solutions, which are both non-circular and non-radial in character. 
\end{abstract}

\begin{keywords}
accretion, accretion discs --- black hole physics 
\end{keywords}
\noindent

\section{introduction} 

The accretion of material onto a black hole is a process of fundamental importance in high energy astrophysics: by ``lighting up'' the black hole's vicinity, accretion offers an unparalleled opportunity to study both  black holes themselves, but also the physics of fluids evolving in the strong gravity regime.  Physical theories describing the accretion process  are  important for interpreting  a broad range of astrophysical observations, but remain not fully understood in all their intricacies.  

Existing accretion solutions were developed in a series of  seminal papers in the 1970s.  This work began with a Newtonian theory of gravity (Lynden-Bell 1969, Shakura and Sunyaev 1973, Lynden-Bell and Pringle 1974) before being extended to the regime of general relativity by Novikov \& Thorne (1973) and Page \& Thorne (1974). These relativistic models form the fundamental basis for the spectral modelling of Galactic X-ray binaries (among other systems), and any model inaccuracies may propagate into potentially substantial theoretical biases in observationally inferred system parameters, such as the source black hole's spin. It is of interest, therefore, to examine any potential shortcomings in these models. 


One of the major limitation of current classical black hole accretion theories pertains to an interesting prediction of the orbital equations of general relativity -- the existence of an innermost stable circular orbit (ISCO) in black hole spacetimes. Within the ISCO, the angular momentum of a circular orbit increases inward, which means that the orbits are unstable: a tiny perturbation from circular motion will eventually acquire a significant inward radial velocity, even while formally conserving its angular momentum and energy. This represents a fundamental change in the behaviour of the flow, from an angular momentum-dominated flow (the outer disc), to a flow no longer constrained by the conservation of its angular momentum (in effect, free-fall). 

Almost all currently used analytical accretion solutions are artificially curtailed at the ISCO  (see Wilkins et al. 2020 for a notable exception, an extension of the surface density through the ISCO). Each thermodynamic (temperature, pressure, density, etc.) quantity is forced to smoothly vanish at the ISCO.   This has the unfortunate additional consequence of forcing other dynamical quantities (such as the flows radial velocity) to diverge.  In reality, if both the radial velocity and mass accretion rate are finite at the ISCO, then all thermodynamic quantities must also be finite at this location. The question then becomes how the fluid evolves throughout the intra-ISCO region, and whether or not the process produces observable emission. 


This is a problem of contemporary astrophysical interest, as observations of black hole systems are beginning to hint at cracks in pre-existing theories. Fabian et al. (2020) argued that observations of the X-ray binary system MAXI J1820 in an extremely bright state could not be successfully modelled  without the addition of {\it ad hoc} thermal emission components with small emitting areas and high emitting temperatures. The characteristics of this additional emission were argued to be similar to what might be expected from the intra-ISCO fluid, a statement which can only be tested with a robust physical theory of this region.  Other observations that may hint at the need for additional physics beyond the standard approach include the ``too-broad'' spectra observed from LMC-X3 by Sutton et al. (2017), and the extreme (beyond the Thorne limit) spins inferred from spectral fitting of the source Cygnus-X1 (Zhao et al. 2021).


The properties of the intra-ISCO accretion fluid has heretofore been studied exclusively by numerical techniques, with a number of general relativistic magneto-hydrodynamical (GRMHD) simulations tailored specifically to the study of this problem (e.g. Schaffe et al. 2008, Noble, Krolik \& Hawley 2010, Penna et al. 2010, Zhu et al. 2012, Schnittman, Krolik \& Noble 2016), as well as numerical extensions of the classical Novikov, Page and Thorne (1973, 1974) equations (Potter 2021, hereafter P21). Every numerical simulation has found that all physical parameters of the accretion flow remain finite across and within the ISCO. 

In this paper we develop a simple analytical theory for this region, which offers significant insight into the properties of fluids evolving in the intra-ISCO regime. Our solutions are carried through in the adiabatic limit, meaning that we neglect the effects of radiative losses and radiative diffusion on the evolving fluid's properties. These profiles must therefore be considered a first step towards a more complete theory of the intra-ISCO region, but which may nevertheless describe with some fidelity real accretion flows in the broad and physically reasonable regime of quasi-adiabatic flows.   



The approach developed in this paper may be thought of as a merging of two classical approaches to studying accretion problems: the Bondi (1952) radial free-fall problem, and the angular momentum dominated flows of accretion discs. Fluid crossing the ISCO has a finite (and large) angular momentum content, but is still able to reach the origin entirely under the influence of gravity, much like the solutions described by a Bondi flow, but now without the need to shed angular momentum.   

In this work we analytically extend, for the first time, relativistic disc solutions through the ISCO, smoothly joining them onto a ``ballistic'' inner flow, creating a global relativistic accretion solution valid at all radii. The smooth joining of these two branches introduces a new and important physical parameter -- $u_I$, the trans-ISCO radial velocity of the disc fluid. The value of $u_I$ is of fundamental importance for understanding the detailed thermodynamic evolution of the intra-ISCO flow.  It is also a parameter, we argue, which is not well-described in current post-ISCO accretion theories.  We emphasise here the need for dedicated calculations to better understand the properties of $u_I$. 


This theory developed here highlights the principal thermodynamic processes at play in this regime. The radial evolution of the disc's central midplane temperature is non-trivial, with radial expansion and vertical compression (both driven by gravitational accelerations) of the flow competing with one another. In fact, we prove the general and remarkable result that these solutions cool for radii $r_I/2 < r < r_I$, before increasing in temperature for $0 < r < r_I/2$, independently of the black hole spin and the assumed equation of state of the accretion flow. Coupled to this behaviour is a plummeting optical depth across the intra-ISCO region, a result of the dropping surface density required by mass conservation. This pronounced drop causes the temperature of the escaping radiation to increase more rapidly than the disc's central temperature.  For some solutions, the disc radiation temperature has its global maximum within the ISCO, and thus the hottest point in the disc may actually be missed in classical disc modelling!


The layout of this paper is the following. In section 2 we recap the theory of the extra-ISCO disc region, highlighting the  assumptions inherent to the modelling which will no longer hold within the ISCO.  In section 3 we derive the dynamic equations of motion (the relativistic Euler equation) of the intra-ISCO accretion fluid, before recapping the work of Mummery \& Balbus (2022) which provide simple analytical solutions of these equations. In section 4 we derive the fundamental thermodynamic evolution equation of the flow, which we solve in the adiabatic limit in section 5. The physical properties of these solutions are discussed in section 6. In section 7 we discuss the validity of the assumption of vertical hydrostatic equilibrium of the intra-ISCO fluid, before concluding in section 8. Some technical results and example Figures are presented in Appendices. 

\section{Classical extra-ISCO theory}
In this section we briefly recap the classical theory of thin relativistic accretion discs, focussing on the behaviour of the fluid near the ISCO, and the assumptions inherent to the theory which break down interior to the ISCO. The reader familiar with this material may wish to skip directly to section 3.  A quantitative treatment of the {extra}-ISCO regime, which we construct so as to smoothly join onto our intra-ISCO solutions, is presented in Appendix \ref{AA}. 

Throughout this paper we will be examining discs evolving in the Kerr metric.    Our notation is as follows.   We use physical units in which we denote  the speed of light by  $c$ and the Newtonian gravitational constant by  $G$.    In coordinates $x^\mu$, the invariant line element ${\rm d}\tau$ is given by
\beq
{\rm d}\tau^2 = - g_{\mu\nu} {\rm d}x^\mu {\rm d}x^\nu,
\eeq
where $g_{\mu\nu}$ is the usual covariant metric tensor with spacetime indices $\mu, \nu$.   {The coordinates are standard $(t, r, \phi, z)$ Boyer-Lindquist in their near-equator form}, where $t$ is time as measured at infinity and the other symbols have their usual quasi-cylindrical interpretation.    We shall work exclusively in the Kerr midplane $\theta =\pi/2$
.   For black hole mass $M$, and angular momentum parameter $a$ (which has dimensions of length and is restricted to $|a| < r_g \equiv GM/c^2$), the non-vanishing $g_{\mu\nu}$ metric components 
 are:
\begin{align}
g_{00} &= -(1 - 2r_g/r)c^2 , \quad g_{0\phi} = g_{\phi0} = -2r_gac/r,  \nonumber \\
g_{\phi\phi} &= r^2+a^2 +2r_ga^2/r, \quad g_{rr} = r^2/\Delta,  \quad g_{zz} = 1, \nonumber \\ 
 \Delta &\equiv r^2 - 2r_g r +a^2, \quad \sqrt{-g} = r, \quad r_g \equiv GM/c^2. 
\end{align}
Note that our choice of coordinates has asymptotic (large radius) metric signature $(-1, +1, +1, +1)$. 

\subsection{Governing disc equations in the stable regime }
The classical theory of extra-ISCO relativistic accretion proceeds by first  defining  a stress energy tensor which describes  the accretion flow and then by constructing mass, energy and momentum conservation equations. The classical accretion disc stress energy tensor is the following
\beq
T^{\mu\nu} = \left(\rho + {P + e \over c^2}\right) u^\mu u^\nu + P g^{\mu\nu} + {1\over c^2}(q^\mu u^\nu + q^\nu u^\mu)
\eeq
where $\rho$ is the rest mass density, $e$ the { internal} energy density and $P$ the pressure of the fluid. The 4-velocity of the flow is $u^\mu$, while $u_\mu$ is its covariant counterpart.  The final pair of terms represent the energy-momentum flux carried out of the system by photons, where $q^\mu$ is the photon flux 4-vector. 
 
With this stress energy tensor defined, one solves the equations of mass, angular momentum and energy conservation 
\beq
\nabla_\mu (\rho u^\mu) = 0 , \quad \nabla_\mu (T^\mu_\phi) = 0, \quad \nabla_\mu (T^\mu_0) = 0, 
\eeq
where in these expressions $\nabla_\mu$ is a covariant derivative with respect to Kerr metric coordinate $x^\mu$. 
These three constraints are sufficient to determine  three key  quantities: the governing equation for the evolution of the disc surface density, the radial velocity of the flow, and the energy flux out of the upper and lower disc surfaces.   

The principal theoretical simplification employed in deriving the thin disc solutions of these coupled equations pertains to the disc fluid velocity. In particular,  the solutions to these three equations are derived by making the following important assumption: the total disc 4-velocity $u^\mu$ (as well as  $u_\mu$) are decomposed into a mean component $U^\mu$ and vanishing-mean fluctuating component $\delta U^\mu$:
\beq
u^\mu = U^\mu + \delta U ^\mu ,\quad u_\mu = U_\mu + \delta U_\mu ,
\eeq
which satisfy asymptotic scalings 
\beq\label{main_scalings}
\delta U_\phi \ll U_\phi, ~~~ U^z \ll U^r \ll \delta U^r \sim \delta U_\phi / r \ll rU^\phi .
\eeq
While the fluctuations are an asymptotic scale larger than the mean radial flow of the disc, they are assumed to vanish on average 
\beq
\left\langle \delta U^\mu \right\rangle \equiv {1 \over \Delta t} \int_t^{t+\Delta t} \delta U^{\mu}(r, t') \, {\rm d}t' = 0,
\eeq
where $\Delta t$ is a time long compared to the timescale upon which turbulent fluctuations are induced in the flow, but much shorter than the timescale upon which the mean disc quantities evolve. While the fluctuations themselves vanish on average, their  average correlations are in general non-zero. In particular, accretion is ultimately driven by the non-zero correlation of the $r-\phi$ components of the turbulent velocity fluctuations, which produce a turbulent stress tensor $W^{\mu\nu}$:
\beq
W^{\mu\nu} \equiv \left\langle \delta U^\mu \delta U^\nu \right\rangle ,
\eeq 
where the angled brackets denote the same averaging procedure introduced above.  As the first order fluctuations in the disc velocity vanish on average, and the second order drift velocity is in general extremely small, the zeroth order motion of the disc is assumed to be that of precisely circular motion, i.e.,  $U^0, U_0, U^\phi$ and $U_\phi$ are equal to the test particle circular motion solutions of the Kerr metric.

Using the above techniques, the following expressions for the energy radiated from the discs upper and lower surfaces:
\begin{equation}\label{energylib}
{\cal F}_{\cal E}=  -   \Sigma W^r_\phi U^0 \,  {{\rm d} \Omega \over {\rm d} r}  ,
\end{equation}
and the mean radial drift velocity of the accretion flow:
\beq\label{radvelgen}
 U^r  =  -  {U^0 \over r \Sigma U_\phi '}  {\partial \over \partial r} \left({ r \Sigma W^r_\phi \over U^0} \right) ,
\eeq
may be derived (Balbus 2017). Here $\Sigma$ is the disc surface density, $W^r_\phi$ is the $r$-$\phi$ component of the turbulent stress tensor, i.e., 
\beq
W^r_\phi \equiv \left\langle \delta U^r \delta U_\phi \right\rangle ,
\eeq 
$U_\phi'$ denotes the specific angular momentum gradient, and $\Omega = U^\phi / U^0$ is the effective angular frequency of the flow. 

From equations \ref{energylib} and \ref{radvelgen} we see that in the main body of the disc the turbulent stress acts {\it both} to drive radial motion via the redistribution of angular momentum  (eq. \ref{radvelgen}), {\it and} to liberate the free energy of the disc shear, resulting in observable emission (eq. \ref{energylib}).  It is important to note however that there  is no fundamental law of physics which requires the turbulence to fulfil both of these roles, and in fact within the ISCO neither of these processes will be driven by turbulent dissipation. Indeed, gravitational dynamics will dominate the radial motion, and while the magnetic fields present in an accretion flow can (and will) redistribute some angular momentum within the intra-ISCO flow, it is not this angular momentum redistribution which drives radial flow: the fluid would be able to plunge across the event horizon, even if it maintained constant angular momentum.  

Furthermore, it is unlikely that any turbulent angular momentum redistribution within the intra-ISCO flow will drive significant turbulent heating. This is because the physical picture associated with equation (\ref{energylib}): a fluid element undergoing several of orbits, slowly liberating its free energy as heat and moving onto another circular orbit at smaller radii, is not an accurate description of the dynamics and energetics of the intra-ISCO flow.  {Instead, we shall assume that any angular momentum redistribution results in negligible heating, and that the emission from the intra-ISCO flow is sourced by a nearly adiabatic, slowly radiating flow, which originates as a hot plasma flowing over the ISCO.}    

{Before we move on to the physics of the intra-ISCO region, it is important to stress the perturbative hierarchy governing the extra-ISCO physical equations, with the radial velocity of the disc at $r \geq r_I$ assumed to be the smallest velocity scale in the problem.}  As can be demonstrated rigorously (a derivation of which we present in Appendix \ref{AA}), the commonly applied vanishing ISCO stress boundary condition results in $U^r(r_I) \rightarrow \infty$, which is both unphysical and in contradiction with the governing assumptions used to derive the model. This result can be most simply understood from mass conservation, which in the steady state reads 
\beq
\dot M = 2\pi r U^r \Sigma = {\rm constant},  
\eeq 
for steady mass flux $\dot M$. A vanishing ISCO stress enforces  $\Sigma(r_I) \rightarrow 0$, as all thermodynamic quantities vanish at the ISCO in this limit.  To enforce a constant $\dot M$, the required solution is a divergent $U^r$. It is obvious that an accretion flow will have both a finite $\dot M$ and $U^r$ at the ISCO radius, and therefore must have finite and non-zero ISCO surface density, along with all other thermodynamic quantities.

Clearly, it will not be possible to connect an outer diffusive flow onto an inner ballistic  flow if the diffusive velocity is infinite at the joining radius!   A reformulation of the flow in the innermost near-ISCO regions of the stable extra-ISCO regime is required.  In Appendix \ref{AA} we demonstrate that any non-zero ISCO stress  results in a finite and non-zero trans-ISCO velocity. Therefore, the simplest resolution of this unphysical behaviour is to consider a disc model with a non-zero ISCO stress.  The presence of a non-zero ISCO stress is also supported by GRMHD simulations of thin accretion flows, as we summarise in the next section.   

\subsection{The ISCO stress }
The notion that magnetic stresses could exert sizeable torques on the inner regions of accretion discs has a long history (Page \& Thorne 1974; Gammie 1999; Krolik 1999). Numerical studies over the past 20 years have generally shown that the magnetic fields that drive the MRI, the process thought to drive angular momentum transfer in discs, lead to extended evolutionary phases with non-zero stresses at the ISCO (e.g. Schaffe et al. 2008, Noble, Krolik \& Hawley 2010, Penna et al. 2010, Zhu et al. 2012, Schnittman, Krolik \& Noble 2016). The principal disagreement which now remains is the dynamical and observational importance of this non-zero ISCO stress, not its existence.

In a disc with a non-zero ISCO stress, angular momentum is transported {outwards} from the unstable disc region ($r < r_I$), back into the stable disc region ($r > r_I$). By the time they reach the event horizon, fluid elements (in numerical simulations) typically have angular momenta 
below what is required of a circular orbit at the ISCO (e.g., a 5--15 percent decrease was found in Noble et al. 2010, while a 2 percent decrease was found in Schaffe et al. 2008). This liberated angular momentum slows the rotation of the inner edge of the stable disc region, and results in the presence of an angular momentum flux from the ISCO neighbourhood. Accretion flows in GRMHD simulations often have their hottest temperatures at the ISCO (e.g., Zhu et al. 2012, Schnittman, Krolik \& Noble 2016).  Previous numerical extensions of the standard thin disc equations through the ISCO (P21) have assumed that the turbulent stress within the ISCO remains well described by an $\alpha$-viscosity. While the modelling of a turbulent stress with an effective viscosity may be a reasonable description in the main body of the disc, it seems unlikely that it will capture the properties of full GRMHD turbulence in the intra-ISCO regime.  It is important to remember that a turbulent flow is not equivalent to a viscous flow, no matter how large the ``coefficient of viscosity'' is made. In certain physical regions, one of which being the intra-ISCO region of a black hole accretion flow, turbulence has fundamentally different properties to an enhanced viscosity, and may be able to transfer angular momentum in physical situations where a viscosity could not.

The solutions of the relativistic thin disc equations with a finite ISCO stress are also hot at the ISCO, with finite values of, for example, the  temperature and density of the flow at this location.  Coupled with a finite non-vanishing trans-ISCO radial velocity, these solutions represent the ideal boundary conditions with which to join onto an intra-ISCO accretion solution. 

The first key physical behaviour to understand within the ISCO is how the small radial drift velocity of the extra-ISCO flow grows into the large plunging velocity of the intra-ISCO region. We turn our attention to this question in the next section.  

\section{Dynamics of the intra-ISCO flow}\label{deriv_dynam}
{As we have emphasised,} in the extra-ISCO `main body' of the disc, there exists a strict velocity hierarchy: the zeroth order behaviour of the flow is that of circular motion, this velocity fluctuates enter at first order, and at second order a radial drift velocity is present. Upon crossing the ISCO this hierarchy will quickly break down, with the radial velocity of the flow becoming of order the speed of light, and very much an `order unity' component of the total flow.  While at first this may appear to be analytically intractable, there is a compensating simplification: the angular momentum and energy of the fluid elements undergoing an intra-ISCO inspiral are unlikely to vary by any significant degree from their ISCO values. This is because turbulent redistribution of these quantities is no longer required to drive radial motion. 

This advection of angular momentum has important consequences for the solution of the relativistic Euler equation of motion, which we now derive. We will {derive in this section a more complicated equation than we end up solving, as it is illuminating to analyse which physical processes are important in this region, which can be neglected, and what the underlying physical basis is for the distinction.} 

\subsection{The relativistic Euler equation}
Consider a fluid described by the following perfect-fluid stress energy tensor, with the additional inclusions of  radiative losses and electromagnetic terms included under the ideal magnetohydrodynamics (MHD) approximation (see e.g., Misner, Thorne and Wheeler 1973) 
\begin{multline}\label{streng}
T^{\mu\nu} = \left(\rho + {P + e \over c^2}\right) U^\mu U^\nu + P g^{\mu\nu} + {1\over c^2} (q^\mu U^\nu + q^\nu U^\mu) \\
+ {1 \over 4\pi \mu_0} \left({1 \over 2} b^2 {U^\mu U^\nu \over c^2} + {1 \over 2} \left({U^\mu U^\nu \over c^2} + g^{\mu\nu}\right) b^2 - b^\mu b^\nu \right) ,
\end{multline}
where again $P$, $e$ and $\rho$ correspond to the total pressure, energy density and (rest) mass density of the fluid  respectively, $U^\mu$ is the fluid's 4-velocity and $q^\mu$ is the photon heat flux 4-vector. {The ideal MHD approximation corresponds to assuming an infinite conductivity of the flow.  In thin accretion discs with moderate accretion rates this is a good assumption, as any electric fields are quickly screened by rearrangements of charged particles. In this expression $b^\mu$ is the magnetic four-vector, defined as 
\beq
b^\mu = {1\over 2} \epsilon^{\mu \nu \lambda \kappa} U_\nu F_{\lambda \kappa} ,
\eeq
where $\epsilon^{\mu\nu\lambda\kappa} = -|g|^{-1/2} \epsilon(\mu\nu\lambda\kappa)$, is the Levi-Civita tensor, note that $\epsilon(\mu\nu\lambda\kappa)$ (the Levi-Civita tensor density) is antisymmetric on all pairs of indices, and $F_{\lambda \kappa}$ is the electromagnetic field (Faraday) tensor. We note that $b^\mu U_\mu = 0$, and that $b^\mu$ corresponds to the magnetic field in the fluid frame $b^\mu_{\rm rest} = (0, \vec B)$. We define the invariant  $b^2 = b_\nu b^\nu$.}
The conservation of energy and momentum of the flow can be compactly described via 
\beq
\nabla_\mu T^{\mu \nu} = 0,
\eeq
a vector equation. Expanding out this equation in full gives 
\begin{multline}
\nabla_\mu T^{\mu \nu} = U^\mu U^\nu {\partial \over \partial x^\mu} \left({P + e \over c^2} + \rho \right) + \left({P + e \over c^2} + \rho \right) U^\nu \nabla_\mu U^\mu \\ + \left({P + e \over c^2} + \rho \right) U^\mu \nabla_\mu U^\nu + g^{\mu \nu} {\partial P \over \partial x^\mu} + {1\over c^2} q^\nu \nabla_\mu U^\mu + {1\over c^2}U^\mu \nabla_\mu q^\nu \\ + {1\over c^2}U^\nu \nabla_\mu q^\mu + {1\over c^2}q^\mu \nabla_\mu U^\nu + {\cal B}^\nu = 0. 
\end{multline}
In this expression we  define the  electromagnetic   contribution as 
\begin{multline}
{\cal B}^\nu \equiv \nabla_\mu T^{\mu\nu}_{\rm mag} = {1 \over 4\pi\mu_0} \Bigg({b^2 \over c^2} U^\mu \nabla_\mu U^\nu + {b^2 \over c^2} U^\nu \nabla_\mu U^\mu  + \\  \left({U^\mu U^\nu \over c^2} + {1\over 2} g^{\mu\nu}\right) \nabla_\mu b^2 - b^\mu\nabla_\mu b^\nu - b^\nu \nabla_\mu b^\mu  \Bigg) .
\end{multline}
Some simplification of this expression can be obtained by noting that 
\beq
 \nabla_\mu U^\mu = - {U^\mu \over \rho} {\partial \rho \over \partial x^\mu}  + {1\over \rho} \nabla_\mu (\rho U^\mu) = - U^\mu {\partial \ln \rho \over \partial x^\mu} ,
 \eeq
 where $\nabla_\mu (\rho U^\mu ) = 0$ by mass conservation. Thus 
 \begin{multline}
 \nabla_\mu T^{\mu \nu} = \left({U^\mu U^\nu \over c^2} + g^{\mu \nu} \right) {\partial P \over \partial x^\mu }  + \left({P + e \over c^2} + \rho \right) U^\mu \nabla_\mu U^\nu  \\
 + {U^\nu \over c^2} \left[ U^\mu {\partial e \over \partial x^\mu} -  {(P + e ) \over \rho } U^\mu  {\partial \rho \over \partial x^\mu} \right] + {1\over c^2} U^\nu \nabla_\mu q^\mu +{1\over c^2}  U^\mu \nabla_\mu q^\nu \\ + {1\over c^2}q^\nu \nabla_\mu U^\mu + {1\over c^2}q^{\mu} \nabla_\mu U^\nu + {\cal B}^\nu = 0. 
\end{multline}
The equation of energy-momentum conservation in this form is rather unwieldy, but it can be further simplified by noting that upon contracting this full expression with $U_\nu$, various terms are equal to zero.  In particular, using the following identities 
\begin{align}
& U^\nu U_\nu = -c^2 ,\\
& U_\nu \nabla_\mu U^\nu =  U^\nu  \nabla_\mu U_\nu =  {1\over 2}  \nabla_\mu (U^\nu U_\nu) = 0,
\end{align}
the top and bottom lines of the above general expression can be shown to be precisely zero upon contraction with $U_\nu$. The two non-trivial terms in this calculation are the expressions for $U_\nu q^\nu$ and $U_\nu {\cal B}^\nu$, both of which vanish identically. The radiative heat flux is given by the 4-vector {(Eckart 1940; see also Chandra et al. 2015 for a discussion of some of the difficulties involved in formulating a relativistic theory of heat flux)}
\beq
q^\mu = -{1 \over \kappa \rho} \left(g^{\mu\nu} + {U^\mu U^\nu \over c^2} \right) \left[\nabla_\nu \left( \sigma T^4 \right) +  \left(U^\gamma {\partial U_\nu \over \partial x^\gamma}\right) {\sigma T^4 \over c^2}  \right]  .
\eeq
{The vanishing of $q^\mu U_\mu$ follows directly as a result of the pre-factor $g^{\mu\nu} + U^\mu U^\nu/c^2$ of the radiative heat flux 4-vector. }

{The formal proof of the precise vanishing of $U_\nu {\cal B}^\nu $ is more involved, although physically it simply corresponds to the well known result that magnetic fields (in the absence of resistivity) do not change the entropy of a fluid. A full proof of $U_\nu {\cal B}^\nu = 0$ is presented in Appendix \ref{mag_proof}, in this section we simply proceed by using the result $U_\nu {\cal B}^\nu = 0$. }

If the top and bottom lines of this expression vanish exactly upon contraction with $U_\nu$, while the middle line does not, it follows that the middle line must by itself independently vanish, and can thus be removed from our equation. This middle line  in fact represent entropy conservation in the flow.  Our governing equation simplifies therefore to 
 \begin{multline}
 \nabla_\mu T^{\mu \nu} = \left({U^\mu U^\nu \over c^2} + g^{\mu \nu} \right) {\partial P \over \partial x^\mu }  + \left({P + e \over c^2} + \rho \right) U^\mu \nabla_\mu U^\nu  \\
 + {1\over c^2} q^\nu \nabla_\mu U^\mu + {1\over c^2} q^{\mu} \nabla_\mu U^\nu + {\cal B}^\nu = 0,
\end{multline}
or equivalently, by using 
\beq
 U^\mu \nabla_\mu U^\nu = U^\mu {\partial U^\nu \over \partial x^\mu } + \Gamma^{\nu}_{\mu \sigma} U^\mu U^\sigma ,
\eeq
where $\Gamma^\alpha_{\beta \gamma}$ is the affine connection of the Kerr spacetime, we find 
 \begin{multline}
U^\mu {\partial U^\nu \over \partial x^\mu }   = -  \Gamma^{\nu}_{\mu \sigma} U^\mu U^\sigma - {c^2 \over P + e + \rho c^2 } \Bigg[ \left({U^\mu U^\nu \over c^2} + g^{\mu \nu} \right) {\partial P \over \partial x^\mu }    \\
 + {1\over c^2} q^\nu \nabla_\mu U^\mu +  {1\over c^2} q^{\mu} \nabla_\mu U^\nu  + {1 \over 4\pi\mu_0} \Bigg({b^2 \over c^2} U^\mu \nabla_\mu U^\nu + {b^2 \over c^2} U^\nu \nabla_\mu U^\mu  + \\  \left({U^\mu U^\nu \over c^2} + {1\over 2} g^{\mu\nu}\right) \nabla_\mu b^2 - b^\mu\nabla_\mu b^\nu - b^\nu \nabla_\mu b^\mu  \Bigg) \Bigg],
\end{multline}
this is the general relativistic analogue of the MHD Euler equation. The terms on the right hand side of this expression each have dimensions of acceleration, and can be loosely identified with the ``acceleration due to gravity'' (first term right hand side), the ``acceleration due to pressure gradients'' (second term), the ``acceleration due to radiation'' (the pair of $q^\mu$ terms on the middle line), and the ``acceleration due to magnetic fields'' (final set of terms). The ``acceleration due to radiation'' of our solutions is generally much smaller than either the gravitational and pressure terms, particularly for the case of the radial momentum equation we are interested in, and will be neglected. 

An important assumption that we shall make going forward is that  the ``acceleration due to gravity" of our solutions, which within the ISCO will be of order  
\beq
a_G \sim \left|\Gamma^r_{\phi\phi} U^\phi U^\phi\right| \sim \left| {GM \over r_I^2} \right| \sim {c^2 \over r_I} ,
\eeq
must be much larger than ``accelerations due to pressure gradients" induced by the trans-ISCO flow 
\beq
a_P \sim \left| {1\over \rho} {{\rm d} P \over {\rm d} r} \right| \sim {c_s^2 \over r_I} \ll a_G \sim {c^2 \over r_I} .
\eeq
This is a robust assumption provided the speed of sound is sub-relativistic. It is simplest to assume that this is the case, and then later check our solutions for consistency with this condition.  We will demonstrate that the solutions constructed in this paper typically satisfy this constraint by many orders of magnitude. 

{Finally, we shall also neglect the explicit contribution of any magnetic fields present on the dynamical evolution of the intra-ISCO flow.  The above equation demonstrates that the magnetic acceleration is of order }
\beq
a_B \sim  {b^\mu \over 4\pi\rho\mu_0 } \nabla_\mu b^\nu \sim {v_A^2 \over r_I} \ll a_G \sim {c^2 \over r_I}.
\eeq
{Provided now that the Alf\'ven velocity is sub-relativistic, this is a good assumption. It is of course possible, perhaps in the so-called ``MAD'' (Magnetically Arrested Disc) state, that this condition may not in fact be satisfied. In such a case the inflow velocity of the disc fluid may deviate from the solutions constructed here, {which would correspond to a very distinct parameter regime.} } 

If we assume that the acceleration due to gravity within the ISCO dominates, then we may simplify the relativistic Euler equation to 
\beq
U^\mu {\partial U^\nu \over \partial x^\mu } \simeq - \Gamma^{\nu }_{\sigma\mu} U^\sigma U^\mu ,
\eeq
which is precisely the geodesic equation for test particle orbits.   Exact solutions of the geodesic equations of the test-particle intra-ISCO inspiral have recently been found by Mummery \& Balbus (2022), which we discuss further below. 

\subsection{The intra-ISCO inspiral }
The leading order solution of the relativistic Euler equation, in the limit of small pressure gradients, is the solution of the geodesic equations for test particle flow. As the  turbulent redistribution of the energy and angular momentum of the accreting fluid is no longer required to drive radial motion,  
we shall employ a `ballistic approximation', and assume that the intra-ISCO flow conserves its angular momentum and energy. These constants of motion are then given by their final values in the stable diffusive regime: the angular momentum $J \equiv U_\phi(r_I)$ and (dimensionless) energy $\gamma \equiv -U_0(r_I)$  of a circular orbit at the ISCO. With this simplification employed we can then derive a simple and universal expression for the inflow velocity of an intra-ISCO inspiral, and from this determine the evolution of the accreting fluids thermodynamic quantities.    

The governing geodesic equation of an intra-ISCO inspiral is easily stated in formal terms:
\beq
g_{\mu\nu} U^\mu U^\nu = g_{rr} (U^r)^2 +U^0U_0 +U^\phi U_\phi =-c^2, 
\eeq
but its direct solution is a matter of no little algebraic complexity.   
After expressing all non-radial 4-velocities in terms of $J$ and $\gamma$ (which are constants of motion), and multiplying through by $1/g_{rr}$, we have
\begin{multline}\label{main_full}
 (U^r)^2 + {J\over r^2} \left( {2r_ga\gamma c \over r} + \left(1 - {2r_g \over r}\right) J \right)\\  - {\gamma c \over r^2} \left[\left(r^2 + a^2 + {2r_ga^2 \over r}\right) \gamma c - {2r_gaJ \over r}\right]  \\= -c^2\left(1+ {a^2\over r^2} - {2r_g\over r} \right).
\end{multline}
This equation formally holds for arbitrary values of $J$ and $\gamma$, but  for the particular values corresponding to an ISCO circular orbit, this equation simplifies dramatically to (Mummery \& Balbus 2022; see also Appendix \ref{ballistic_appendix}): 
\beq\label{flow_soln}
U^r = - c \sqrt{2r_g \over 3 r_I} \left( {r_I \over r} - 1\right)^{3/2} .
\eeq
See Mummery \& Balbus (2022) for a much more detailed treatment, and a number of explicit $r(\phi)$ solutions, of the intra-ISCO inspiral.  This potentially surprising simplification can be understood by noting the following: equation \ref{main_full} is of the form $(U^r)^2 + V_{\rm eff}(r) = 0$, which defines an `effective' potential $V_{\rm eff}$ which is cubic in $1/r$.   For a circular orbit of radius $r=r_c$,  both $V_{\rm eff}(r_c) = 0$ and $\partial_r V_{\rm eff}(r_c) = 0$, and there will be a double root of the polynomial.    For the particular case of the last stable circular orbit, there is an additional condition, $\partial^2_r V_{\rm eff}(r_I) = 0$.  Thus,  $r_I$ is a {\it triple} root of $U^r$. The normalisation can then be found by taking the formal $r \rightarrow \infty$ limit of both expressions. This expression will form the basis of our intra-ISCO accretion solutions. 

In a real accretion flow, the fluid which crosses the ISCO will have a non-zero radial velocity $u_I$,  and so the angular momentum and energy of the fluid particles will not be precisely those  of a circular orbit at the ISCO. In addition, the angular momentum and energy of the fluid will not be precisely conserved over the inspiral (some angular momentum is passed back to the disc as the source of the ISCO stress). This will cause small modifications to the inflow velocity of the fluid over its inspiral from the ISCO. As the small modifications to the angular momentum and energy of the fluid elements must conspire to produce a radial velocity of $u_I$ at the ISCO, we note that the radial velocity of the intra-ISCO flow will be well approximated by the following expression 
\beq\label{flow_eq}
U^r \simeq - c \sqrt{2r_g \over 3 r_I} \left( {r_I \over r} - 1\right)^{3/2}  - u_I .
\eeq
{We note that a velocity profile of this form is supported by the results of full GRMHD simulations. Schnittman et al. (2016) compared equation (\ref{main_full}), which of course simplifies to eq. (\ref{flow_soln}), to the  intra-ISCO radial velocity found in their simulations (their Fig. 5), finding good agreement.    }

{The assumption of pure geodesic motion is of course a simplification of the complex and turbulent motion of ``real'' accretion flows. We remind the reader that this simplification is premised on the  sound and Alfv\'en velocities of the flow being sub-relativistic, which should be robust for the vast majority of parameter space. However, in the small radial region surrounding the ISCO, where the fluid elements undergo many orbits (Mummery \& Balbus 2022), non-axisymmetric turbulent perturbations can lead to intersections between different inspiralling trajectories, which could act to heat the flow. (Note that pure geodesic inspirals do not intercept, as we prove in Appendix D.) This may be particularly important for high pro-grade spins, where the ISCO and event horizon are located closer together. }  

With the radial velocity of the disc pre-specified, the accretion equations simplify dramatically. The constant mass accretion rate through the disc is given by 
\beq\label{Ssol}
\dot M =  2\pi r \Sigma U^r = {\rm constant}, 
\eeq
thus for $r \leq r_I$ the surface density evolves as 
\beq
\Sigma(r) = \Sigma(r_I) \left({r_I \over r}\right)  \left({u_I \over \left| U^r \right| }\right) .
\eeq
Clearly, the surface density of the disc will fall rapidly over the intra-ISCO inspiral. This will lead to two competing physical changes within the disc flow: (1) the discs central temperature will drop as the disc fluid is stretched radially along the inflow, and (2) the discs optical depth will drop, allowing the discs radiation field to more easily escape. Additionally, the increasing vertical gravity of the near-event horizon regions will compress the vertical scale height of the disc, acting to heat the disc.  These competing effects enter the fluids thermodynamic equation, which we now derive.

\section{The  thermodynamic equation of the intra-ISCO flow}\label{deriv_therm}
The analysis of the proceeding section focussed on the dynamical evolution of the intra-ISCO accretion flow. In this section we focus on the fluids governing thermodynamic equation. The expressions for the fluids energy-momentum conservation can, as we have shown, be written as
 \begin{multline}
 \nabla_\mu T^{\mu \nu} = \left({U^\mu U^\nu \over c^2} + g^{\mu \nu} \right) {\partial P \over \partial x^\mu }  + \left({P + e \over c^2} + \rho \right) U^\mu \nabla_\mu U^\nu  \\
 + {U^\nu \over c^2} \left[ U^\mu {\partial e \over \partial x^\mu} -  {(P + e ) \over \rho } U^\mu  {\partial \rho \over \partial x^\mu} \right] + {1\over c^2} U^\nu \nabla_\mu q^\mu +{1\over c^2}  U^\mu \nabla_\mu q^\nu \\ + {1\over c^2}q^\nu \nabla_\mu U^\mu + {1\over c^2}q^{\mu} \nabla_\mu U^\nu + {\cal B}^\nu  = 0.
\end{multline}
To construct a thermodynamic entropy equation from this expression it suffices to contract this equation with $U_\nu$. As we demonstrated in the proceeding section the first and third lines of this equation vanish identically following this contraction, and we are left with 
 \begin{multline}
 U_\nu \nabla_\mu T^{\mu \nu} =  {U_\nu U^\nu \over c^2} \left[ U^\mu {\partial e \over \partial x^\mu} -  {(P + e ) \over \rho } U^\mu  {\partial \rho \over \partial x^\mu} \right] \\ + {U_\nu U^\nu \over c^2} \nabla_\mu q^\mu + {U_\nu U^\mu \over c^2} \nabla_\mu q^\nu = 0. 
\end{multline}
Simplifying with $U_\nu U^\nu = -c^2$ leaves 
\begin{equation}\label{entropy_cons_simp}
{{\rm d} e \over {\rm d} \tau } - \left(P + e \right) {{\rm d} \ln \rho \over {\rm d} \tau}  = -\left(\nabla_\mu - {1\over c^2} U^\nu U_\mu \nabla_\nu \right) q^\mu . 
\end{equation}
In this expression $\tau$ is the fluid element's proper time
\beq
{{\rm d} \over {\rm d} \tau } \equiv U^\mu {\partial \over \partial x^\mu} .  
\eeq 
For an azimuthally symmetric and geometrically  thin flow, we may further assume that $\partial_\phi \equiv 0$ and $U^z \equiv 0$, and therefore in the steady state ($\partial_t \equiv 0$)
\beq
{{\rm d} \over {\rm d} \tau } \equiv U^r {{\rm d} \over {\rm d} r} .  
\eeq

Equation (\ref{entropy_cons_simp}) describes the conservation of entropy of the intra-ISCO accretion flow.  The left hand side of this expression represents the change in entropy of the disc fluid which, assuming there is no turbulent heating of the intra-ISCO flow, is given only by the flux of heat out of the disc system carried by the photon field (the right hand side of this equation). 

Equation (\ref{entropy_cons_simp}) is the governing equation describing the thermodynamic evolution of the intra-ISCO fluid. A full numerical exploration of the solutions of eq. (\ref{entropy_cons_simp}) shall be presented in a follow-up work, but for present purposes we restrict our analysis to a simplified, but physically well-motivated,  analytically tractable limit. 

The character of the solutions of equation (\ref{entropy_cons_simp}) are effectively determined by whichever of two physical processes dominates:  adiabatic or radiative flow.  The adiabatic terms are described by the left hand side of eq. (\ref{entropy_cons_simp}) (those terms associated with the stretching and compression of the fluid's volume).  These terms evolve on the free-fall timescale, roughly given by 
\beq
t_{\rm ff} \sim r/|U^r| . 
\eeq  
For intra-ISCO evolution, this is an extremely short timescale, not much longer than the light crossing time at the ISCO. 

The right hand side of eq. (\ref{entropy_cons_simp}) concerns the evolution driven by radiative processes. The interplay of the various radiative terms on the evolution is generally non-trivial, as it involves both energy-loss terms describing the photon flux out of the disc surface ($\nabla_z q^z$), and the radiative diffusion of photons from the stable outer disc into the intra-ISCO region ($\nabla_r q^r$). These two terms work against each other, with the $\nabla_z q^z$ term cooling the flow (removing entropy) while the $\nabla_r q^r$ term heats the flow (entropy inflow from the outer disc). The interplay of these two terms is best understood via a numerical analysis. 

For the present, we note that in the absence of radiative losses $\nabla_\mu q^\mu \simeq 0$, the evolution equation becomes significantly more analytically tractable.  Physically, this limit corresponds to an adiabatic intra-ISCO evolution, or equivalently the (physically relevant) limit of an extremely short free-fall timescale.  In this limit our governing thermodynamic equation is 
\begin{equation}\label{entropy_cons}
{{\rm d} e \over {\rm d} \tau } - \left(P + e \right) {{\rm d} \ln \rho \over {\rm d} \tau}  = 0. 
\end{equation}
Equation (\ref{entropy_cons})  expresses the conservation of entropy of an adiabatically evolving accretion flow, once it has passed beyond the ISCO.

\section{Analytical  solutions of the adiabatic intra-ISCO accretion problem}\label{iscobondi}
We consider accretion flows where the energy density $e$ and pressure $P$  are related by the following simple equation of state
\beq
e = {P \over \Gamma - 1}, 
\eeq
which is an excellent approximation in regions where the nonrelativistic gas pressure ($\Gamma_g = 5/3$) dominates, or in regions when the radiation pressure ($\Gamma_r = 4/3$) dominates.    Under these conditions, conservation of the fluid's entropy (eq. \ref{entropy_cons_simp}) implies 
\beq\label{Psol}
{P \over \Gamma - 1} {{\rm d} \over {\rm d} \tau} \left(\ln P\rho^{-\Gamma}\right) = 0\quad  \rightarrow \quad P\rho^{-\Gamma} = {\rm constant} .
\eeq 
We will keep the expression general throughout this section, leaving $\Gamma>1$ as a parameter in the problem. 
The density of the flow is related to the discs surface density and scale height $H$ trivially through 
\beq\label{rhsol}
\rho \equiv {\Sigma \over H} .
\eeq
The discs scale height is given by the constrains of vertical hydrostatic equilibrium. Abramowicz {\it et al}. (1997) derived the solution to these constraints, finding a relationship which holds down to the event horizon with corrections at the level of ${\cal O}(H/r)^4 $ 
\beq
H = \sqrt{P r^4 \over \rho (U_\phi^2 + a^2 c^2 (1 - U_0^2))}  .
\eeq
 Within the ISCO $U_\phi$ and $U_0$ are of course constants of motion, given by their values at the ISCO (this is the geodesic dynamics assumption). In fact, for the precise ISCO values  (which we denote $U_\phi = J$ and $U_0 = -\gamma$ to match the notation of eq. \ref{main_full}) this combination simplifies dramatically. This can be seen by noting that the above combination is equal to the coefficient of the $1/r^2$ term of the flow equation (eq. \ref{main_full}). This means that it can be computed by expanding equation \ref{flow_soln}, and noting the corresponding coefficient of the $1/r^2$ term:
\begin{multline}
(U^r)^2 -  {2r_g\over 3r_I} c^2 \left({r_I \over r} - 1\right)^{3} \\= (U^r)^2 - {2r_g r_I^2 \over 3 r^3} c^2 + {2r_gr_I \over r^2}c^2 - {2r_g \over r}c^2 + {2 r_g \over 3 r_I}c^2 = 0,
\end{multline}
in other words
\beq
J^2 + c^2a^2(1 - \gamma^2) = 2GM r_I,
\eeq
 for any value of the black hole spin. We therefore find 
\beq\label{hsol}
H = \sqrt{ {P  \over \rho } {r^4 \over 2GMr_I}}  .
\eeq
The intra-ISCO accretion flow will remain in vertical hydrostatic equilibrium provided that the velocity required to maintain that equilibrium (which we shall identify as $U^z$) remains sub-sonic 
\beq
U^z \leq c_s . 
\eeq
The velocity required to maintain vertical hydrostatic equilibrium of the flow is given by 
\beq
U^z = {{\rm d} H \over {\rm d} \tau} = U^r {\partial H \over \partial r} .
\eeq
From the above expression for the disc scale height, and noting that 
\beq
c_s = \sqrt{{\rm d} P \over {\rm d \rho}} = \sqrt{\Gamma P \over \rho},
\eeq 
we have
\beq\label{trans_sonic_condition}
{U^z \over c_s } = {1\over \sqrt{\Gamma}} {U^r \over c} \left[ 2 +  {\partial \ln c_s \over \partial \ln r}  \right] \sqrt{r^2 \over 2 r_g r_I} ,
\eeq
and thus near to the ISCO (where $U^r \ll c$) vertical hydrostatic equilibrium will be maintained. Closer to the event horizon the validity of this expression will be tested directly from the final solutions. It will be shown that the velocity required to maintain vertical hydrostatic equilibrium will be sub-sonic at all radii exterior to the black hole's event horizon, except for extreme retrograde black hole spins in a region close to the event horizon. For these systems the assumption of strict vertical hydrostatic equilibrium must be revisited, and we will discuss the solutions in that regime in a later section.

We  have four equations linking $\Sigma$, $P$, $\rho$ and $H$ to $U^r$ and $r$, which are both known functions.   Solving these equations (\ref{Ssol}, \ref{Psol}, \ref{rhsol} and \ref{hsol}) in full we find 
\begin{align}
{P \over P_I} &=  \left({r_I \over r} \right)^{6\Gamma/(\Gamma + 1)} \left[ \varepsilon^{-1} \left({r_I \over r} - 1\right)^{3/2} + 1\right]^{-2\Gamma/(\Gamma + 1)} , \\
{\rho \over \rho_I} &= \left({r_I \over r} \right)^{6/(\Gamma + 1)} \left[ \varepsilon^{-1} \left({r_I \over r} - 1\right)^{3/2} + 1\right]^{-2/(\Gamma + 1)} , \label{rhrh}\\
{H \over H_I} &= \left({r_I \over r} \right)^{-(5-\Gamma)/(\Gamma + 1)} \left[ \varepsilon^{-1} \left({r_I \over r} - 1\right)^{3/2} + 1\right]^{-(\Gamma - 1)/(\Gamma + 1)} , \\
{\Sigma \over \Sigma_I} &= \left({r_I \over r} \right) \left[ \varepsilon^{-1} \left({r_I \over r} - 1\right)^{3/2} + 1\right]^{-1} ,
\end{align}
 In these expressions $\varepsilon$ has the definition 
 \beq
 \varepsilon \equiv {u_I \over c} \sqrt{3 r_I \over 2 r_g} \ll 1 ,
 \eeq 
 and each variable with subscript $I$ corresponds to the value of that variable at the ISCO. This is the key result of this paper. {Note that the quantity in the square brackets above is equal to the velocity ratio  }
 \beq
{ |U^r | \over u_I} =  \left[ \varepsilon^{-1} \left({r_I \over r} - 1\right)^{3/2} + 1\right]
 \eeq
 
One of the remarkable properties of these solutions is that they are self-similar, and depend only on $\varepsilon$ and the radius {\it in units of the ISCO} $r/r_I$. {The finding of self-similar profiles of the thermodynamic disc properties is extremely suggestive, as it is supported by the numerical simulations  of Schnittman et al. (2016) who found a ``simple, universal, emissivity profile'', which depended only on the ratio $\tilde r / r_I$.  (Here, $\tilde r$ is the proper radial distance, not radial metric coordinate, a distinction which is unimportant close to the ISCO but grows in importance when the event horizon is approached).  } Future detailed comparisons between the solutions derived here and the results of GRMHD simulations will be of great interest. 
 
 Just as in the case of the extra-ISCO accretion solutions of Novikov \& Thorne (1974) and Page \& Thorne (1974), the precise dependences of the central temperature and related thermodynamic quantities depends on the dominant balance of the radiation and gas pressures within the flow, and on the principal  source of opacity within the disc.   We now discuss three limiting cases below.

\subsection{ { { Gas pressure supported systems, $P_g \gg P_r$ }}}
In the first instance we shall assume that the gas pressure is  dominant, a reasonable assumption for most parameter regimes as the radiation pressure term $P_r \propto T_c^4$ will rapidly drop as the disc fluid adiabatically cools. Thus, we have 
\beq
P \simeq P_g = {k_B \rho T_c \over \mu m_p} , \quad e \simeq {3 \over 2} P_g,  \quad \Gamma = {5 \over 3}.
\eeq
and we can solve for the central temperature evolution.  The explicit solutions for $\Gamma = 5/3$, and gas pressure dominating radiation pressure, are the following: 
 \begin{align}
T_c &= T_{c, I}  \left({r_I \over r}\right)^{3/2}  \left[{\varepsilon}^{-1}\left({r_I \over r} - 1 \right)^{3/2} + 1 \right]^{-1/2} , \label{Tcgas} \\
P_g &= P_{g, I}   \left({r_I \over r}\right)^{15/4}  \left[{\varepsilon}^{-1}\left({r_I \over r} - 1 \right)^{3/2} + 1 \right]^{-5/4} , \\
P_r &= P_{r,I}   \left({r_I \over r}\right)^{6}  \left[{\varepsilon}^{-1}\left({r_I \over r} - 1 \right)^{3/2} + 1 \right]^{-2}  . \label{Prgas} 
\end{align}
 
In deriving these expression we have neglected changes in the discs entropy due to its emission of radiation. However, the disc will of course continue to radiate between $r_I$ and the event horizon ($r_H$), and we can `post-process' the radiation resulting from our central temperature profile, working under the assumption that this radiation will not significantly modify the disc temperature profile, while bearing in mind the potential limitations of this approach.   (The neglect of radiation will likely breakdown for the extremal regions of the parameter space.)   Provided the disc remains optically thick $\kappa \Sigma > 1$ (a condition that can be checked simply for our flow solutions), the radiative temperature may be computed from this central temperature by 
\beq\label{opt_thick}
T_R^4 \propto T_c^4 / \kappa \Sigma. 
\eeq 
If constant electron scattering opacity dominates throughout the plunging region, this would imply
\beq
T_R = T_{R, I} \left[{T_c \over  T_{c, I}}\right] \left[{\Sigma \over \Sigma_I } \right]^{-1/4} ,  \quad \tau = \tau_I  \left[{\Sigma \over \Sigma_I } \right]
\eeq 
or explicitly:
\begin{align}
\tau &= \tau_I  \left({r_I \over r}\right)  \left[{\varepsilon}^{-1}\left({r_I \over r} - 1 \right)^{3/2} + 1 \right]^{-1} , \\
T_R &= T_{R, I}   \left({r_I \over r}\right)^{5/4}  \left[{\varepsilon}^{-1}\left({r_I \over r} - 1 \right)^{3/2} + 1 \right]^{-1/4} .
\end{align}
The assumption that the electron scattering opacity dominates the free-free opacity is, for most black hole disc systems, rather good throughout the main body of the disc. However, it is interesting to note that the free-free Kramers opacity, which depends strongly on disc density and temperature, will grow throughout the plunging region:
 \beq
 \kappa_{\rm ff}  = \kappa_0 \rho T_c^{-7/2}  = \kappa_{{\rm ff}, I}  \left({r_I \over r}\right)^{-3}  \left[{\varepsilon}^{-1}\left({r_I \over r} - 1 \right)^{3/2} + 1 \right] .
 \eeq
(In the first equality  $\kappa_0$ is a constant.) 
For relatively cool discs, where $\kappa_{{\rm ff}, I} \sim \kappa_{\rm es}$, free-free opacity will dominate over the electron scattering opacity  for the entire intra-ISCO regime, and we find the free-free asymptotic solution of the optical depth:
 \beq
 \tau = \tau_I \left({r_I \over r} \right)^{-2} , \quad \kappa_{{\rm ff}, I} \gg \kappa_{\rm es} ,
 \eeq
which falls slowly throughout the intra-ISCO region, despite the much more rapidly dropping disc surface density.  The radiative temperature in this limit evolves as 
 \beq
 T_R = T_{R, I}  \left({r_I \over r}\right)^{2}  \left[{\varepsilon}^{-1}\left({r_I \over r} - 1 \right)^{3/2} + 1 \right]^{-1/2} , \quad \kappa_{{\rm ff}, I} \gg \kappa_{\rm es}.
 \eeq
In the cool-disc $\kappa_{{\rm ff}, I} \sim \kappa_{\rm es}$ limit, we find a steeper fall off of the radiative temperature across the plunging region.

\begin{figure}
\centering
  \includegraphics[width=.95\linewidth]{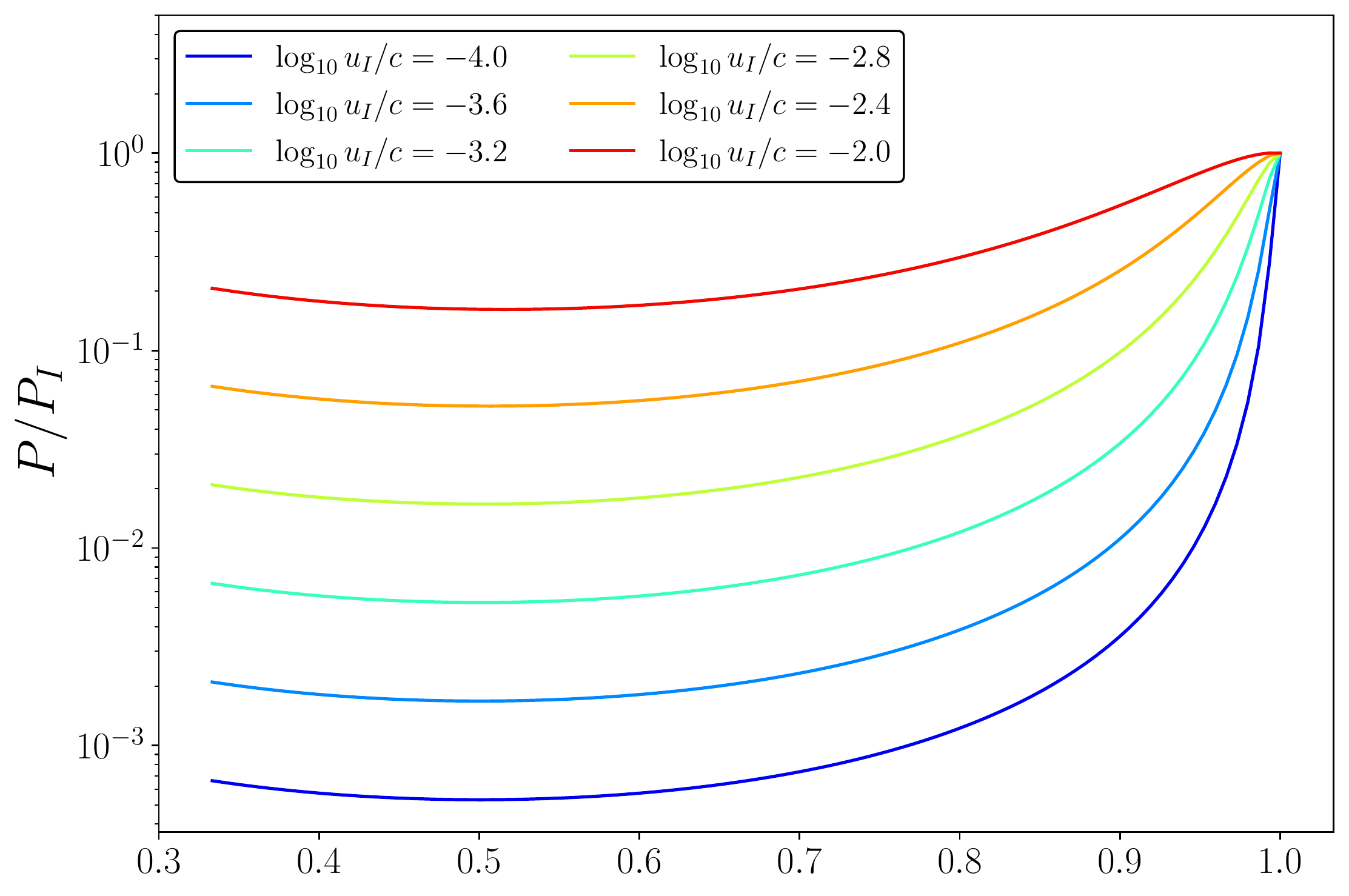} 
  \includegraphics[width=.95\linewidth]{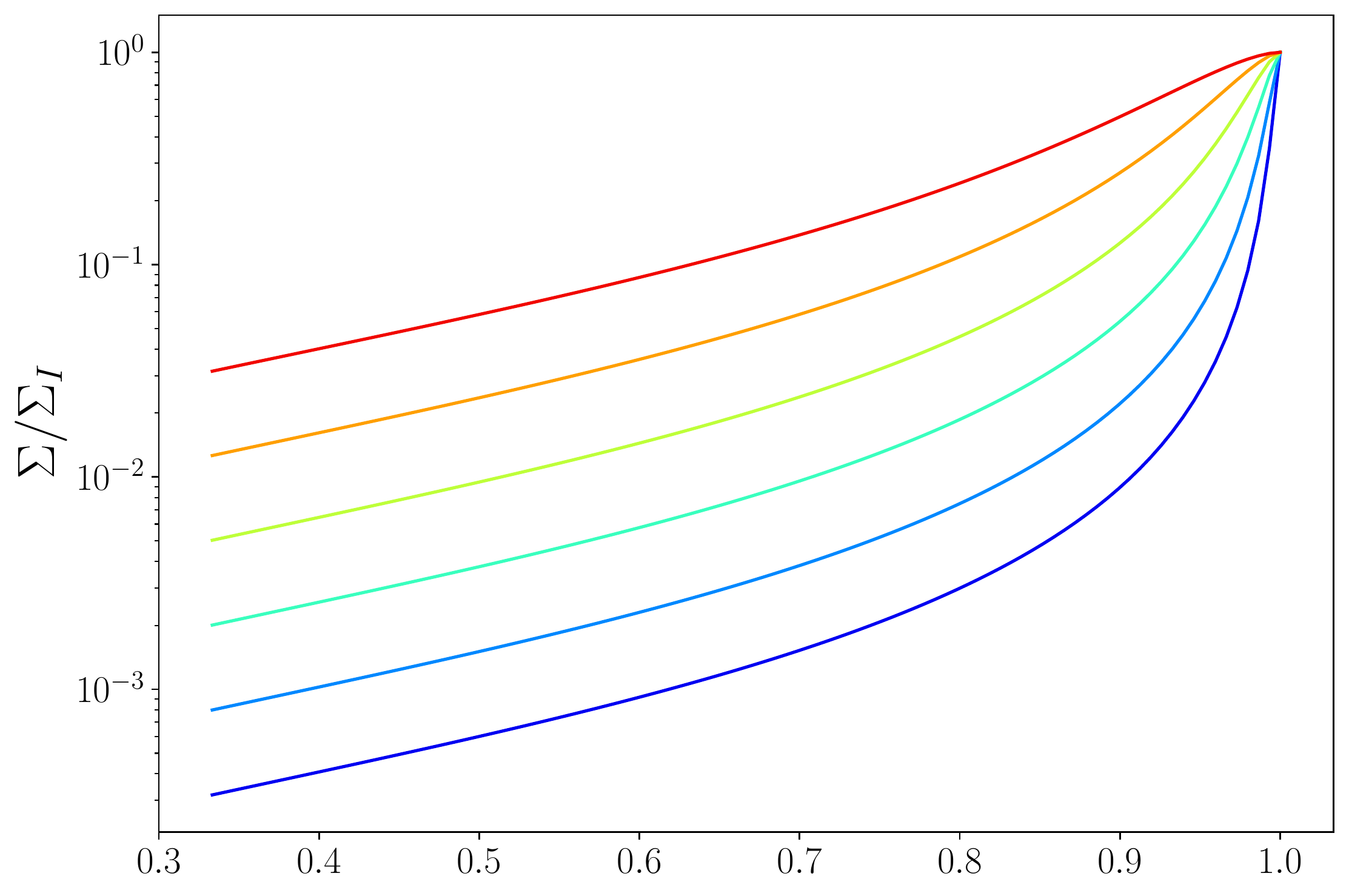} 
    \includegraphics[width=.95\linewidth]{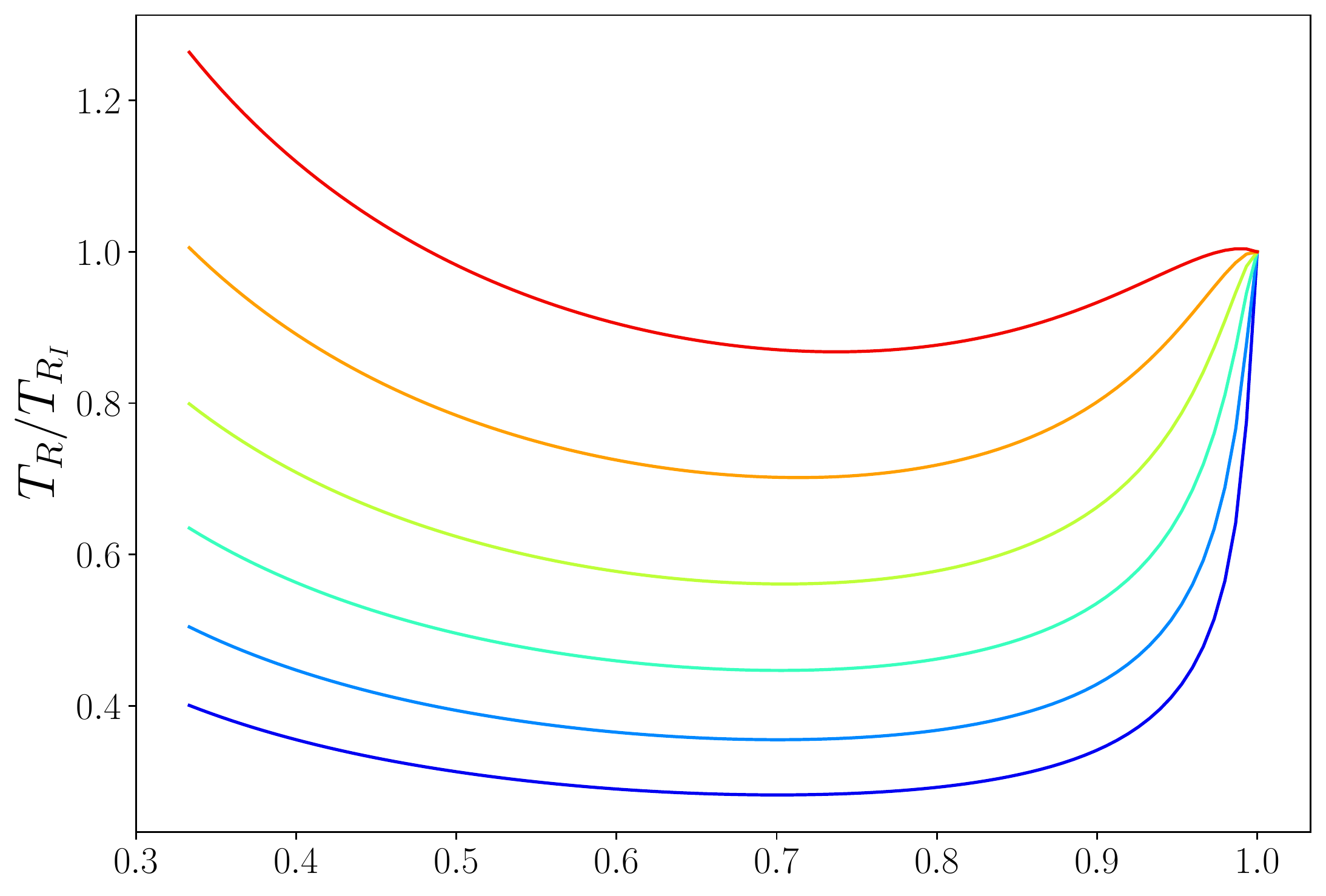} 
    \includegraphics[width=.95\linewidth]{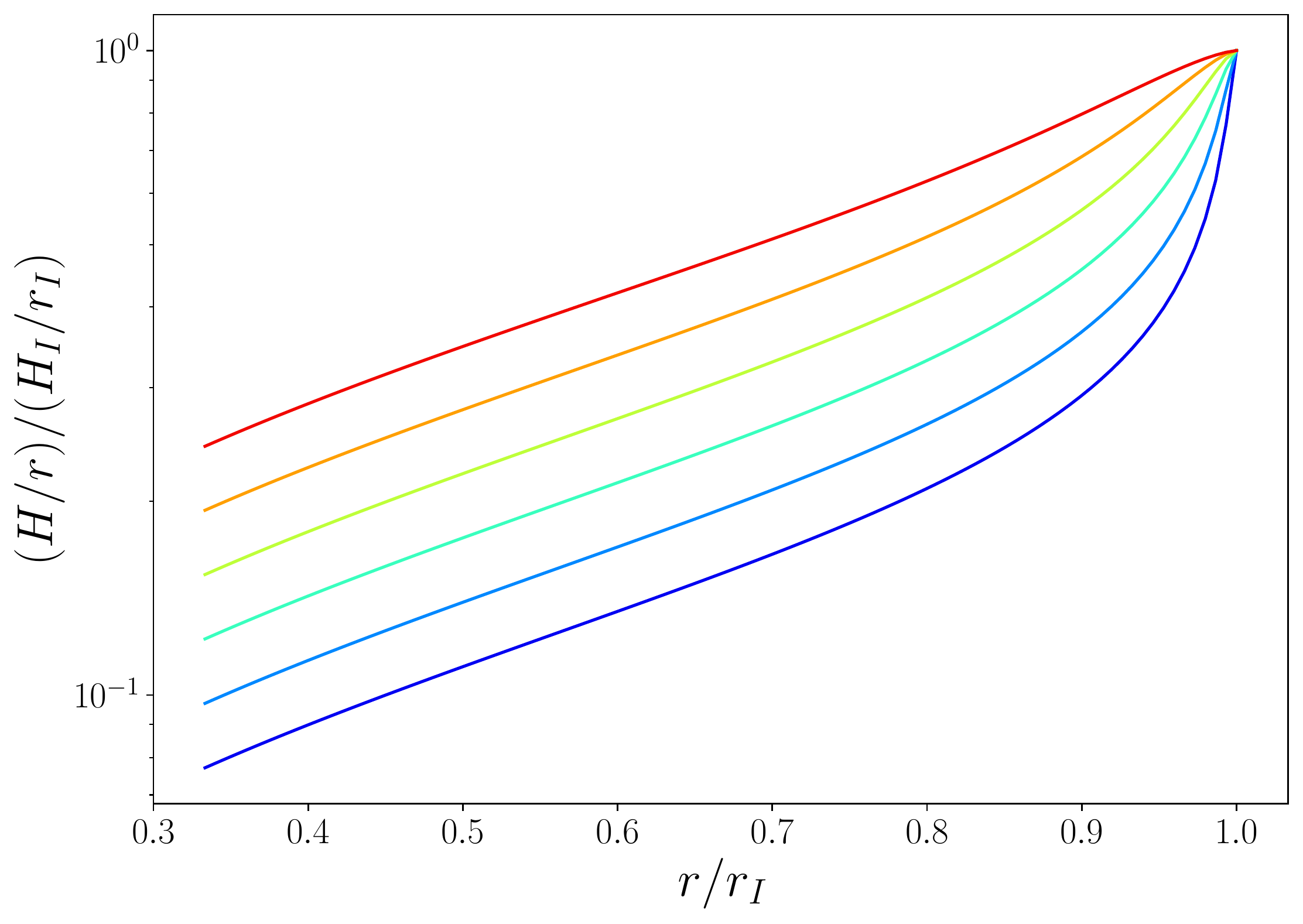} 
     \caption{The total pressure $P$, the surface density $\Sigma$, the radiative temperature $T_R$ and the disc aspect ratio $H/r$, normalised by their ISCO values,  throughout the intra-ISCO region for a number of different trans-ISCO velocities $u_I$. Note the linear vertical scale on the $T_R$ plot, in  contrast with the logarithmic vertical axis of the other plots. This Figure was produced with $\Gamma = 5/3$ (i.e., a gas pressure dominated flow).   } 
 \label{fig1}
\end{figure}

 \subsection{ { { Radiation pressure supported systems, $P_r \gg P_g$ }}}
If the radiation pressure dominates over the gas pressure, we have the following simplification: 
\beq
P \simeq P_r = {4 \sigma T_c^4 \over 3 c} , \quad e \simeq 3 P_r,  \quad \Gamma = {4 \over 3}.
\eeq
leading to
\begin{align}
T_c &= T_{c, I}  \left({r_I \over r}\right)^{6/7}  \left[{\varepsilon}^{-1}\left({r_I \over r} - 1 \right)^{3/2} + 1 \right]^{-2/7} , \label{Tcrad} \\ 
P_g&= P_{g, I}  \left({r_I \over r}\right)^{24/7}  \left[{\varepsilon}^{-1}\left({r_I \over r} - 1 \right)^{3/2} + 1 \right]^{-8/7} , \\ 
P_r&= P_{r, I}  \left({r_I \over r}\right)^{24/7}  \left[{\varepsilon}^{-1}\left({r_I \over r} - 1 \right)^{3/2} + 1 \right]^{-8/7} . \label{Prrad}  
\end{align}
Note that for radiation pressure dominated flows,
\beq
{P_r(r<r_I) \over P_g(r<r_I)} = {P_{r, I} \over P_{g, I}} = {\rm constant} . 
\eeq
The Kramers free-free opacity now grows very slowly across the intra-ISCO region:
\beq
\kappa_{\rm ff} =  \kappa_{{\rm ff}, I} \left({r_I \over r}\right)^{-3/7}  \left[{\varepsilon}^{-1}\left({r_I \over r} - 1 \right)^{3/2} + 1 \right]^{1/7} ,
\eeq
When radiation pressure dominates over gas pressure, we expect $\kappa_{\rm ff}(r_I) \ll \kappa_{\rm es}$ (the free-free opacity only dominates over the electron scattering opacity at low temperatures and high densities, the exact opposite limit to when the radiation pressure dominates over the gas pressure).  Since the free-free opacity only then grows very slowly throughout the intra-ISCO region,  we may neglect its contribution to the optical depth, as such the radiative temperature satisfies 
\beq
T_R = T_{R, I}   \left({r_I \over r}\right)^{17/28}  \left[{\varepsilon}^{-1}\left({r_I \over r} - 1 \right)^{3/2} + 1 \right]^{-1/28} .
\eeq
It is interesting to note that radiation supported discs remain hotter (relative to their ISCO temperature) than those supported by gas pressure. This result is an example of a general principle of these solutions: {\it hotter discs remain hot, while cooler discs cool most rapidly}. This can be seen by examining how the radiative temperature falls over the plunging region (from ISCO to event horizon $r_H$)  in the three governing limits of the problem:
\begin{align}
T_R(r_H) / T_R(r_I) &\sim \varepsilon^{1/2}, \quad \,\,\, \kappa_{\rm ff} \gg \kappa_{\rm es}, \quad P_g \gg P_r, \\
T_R(r_H) / T_R(r_I) &\sim \varepsilon^{1/4}, \quad \,\,\, \kappa_{\rm ff} \ll \kappa_{\rm es}, \quad P_g \gg P_r, \\
T_R(r_H) / T_R(r_I) &\sim \varepsilon^{1/28}, \quad \kappa_{\rm ff} \ll \kappa_{\rm es}, \quad P_g \ll P_r. 
\end{align}
These solutions are presented in order of increasing temperature at the ISCO, and it may be seen that this is also in order of decreasing change in radiative temperature.

\subsection{Example  solutions}

\begin{figure}
\centering
  \includegraphics[width=\linewidth]{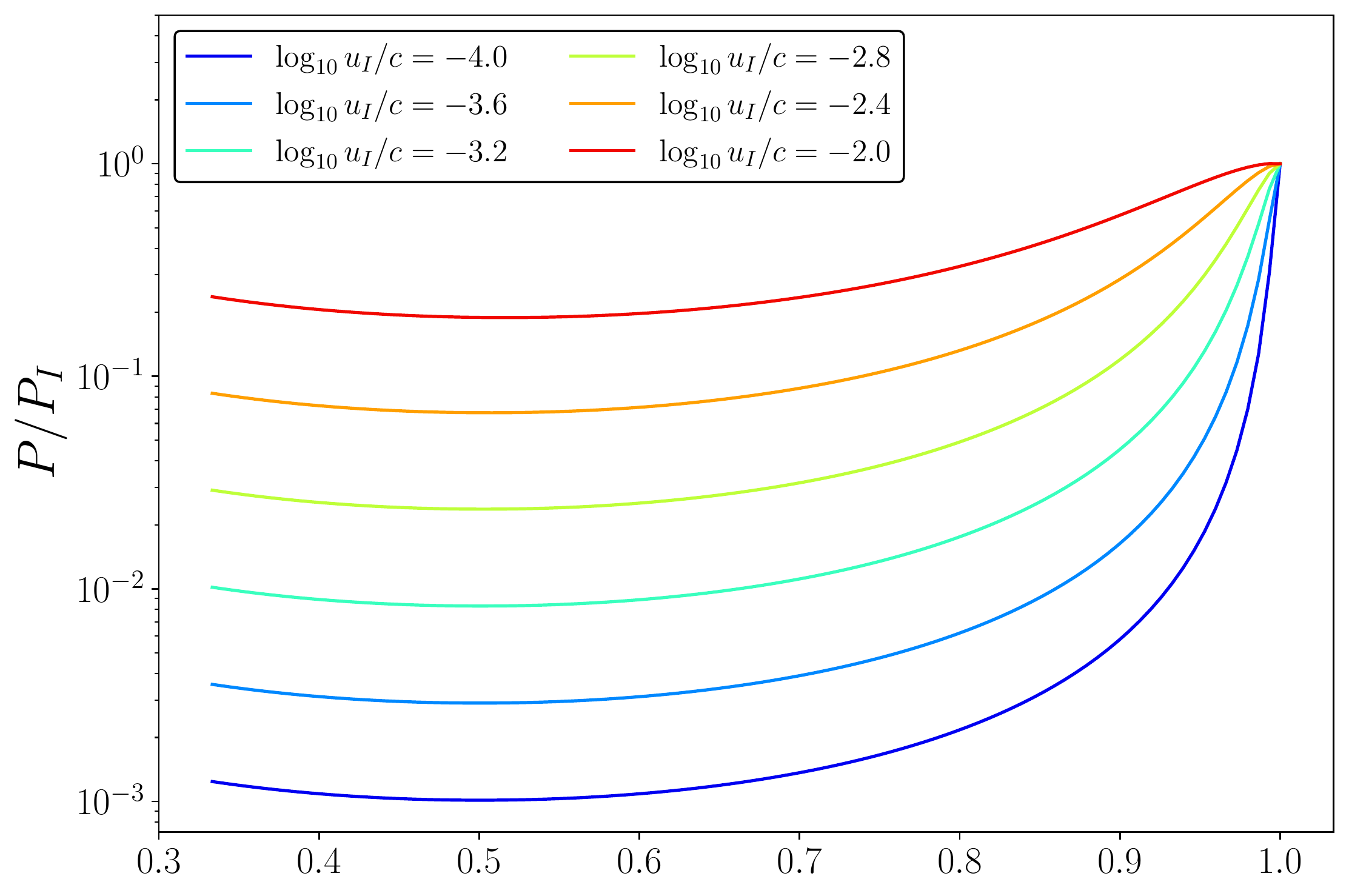} 
  \includegraphics[width=\linewidth]{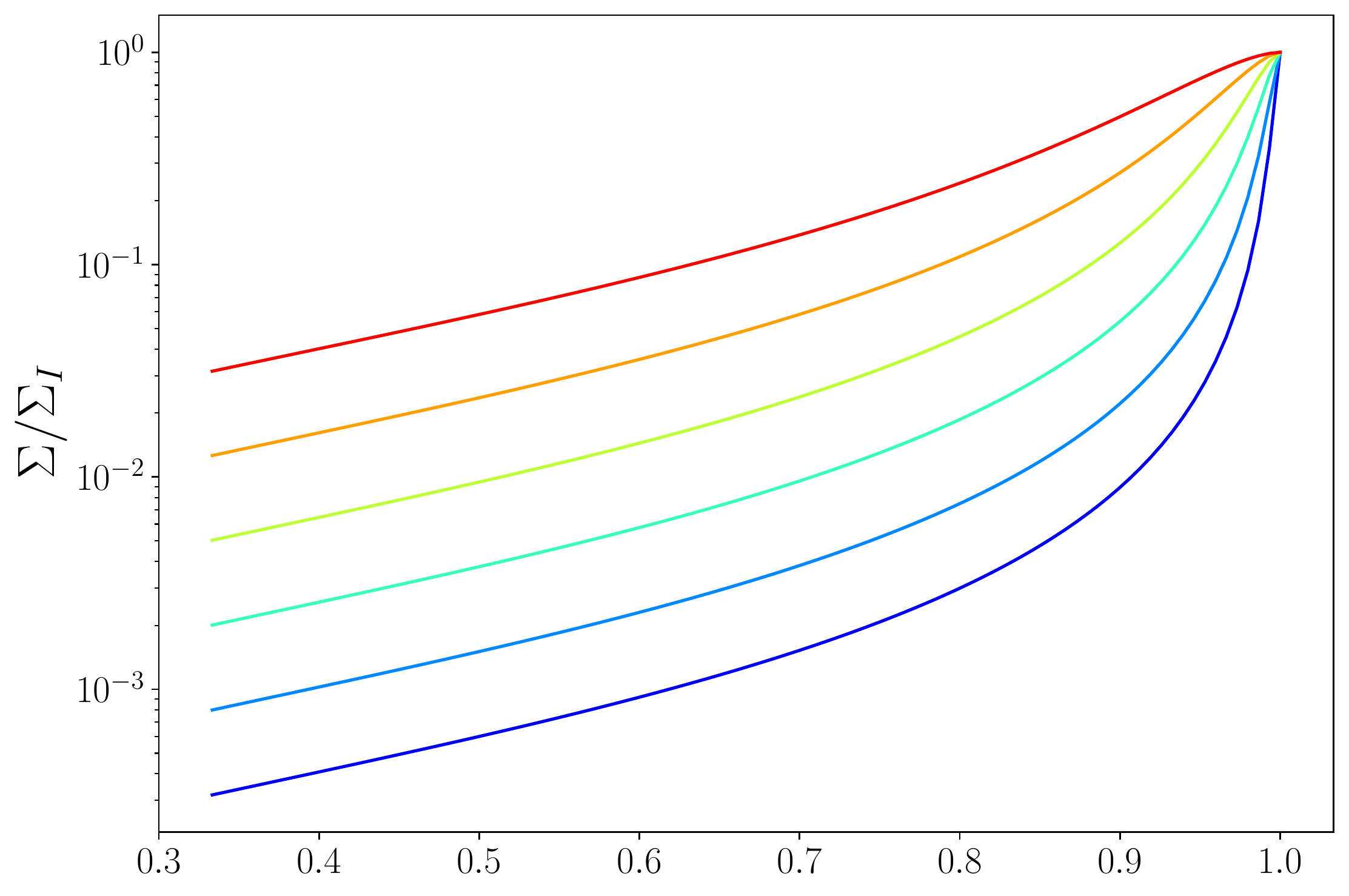} 
    \includegraphics[width=\linewidth]{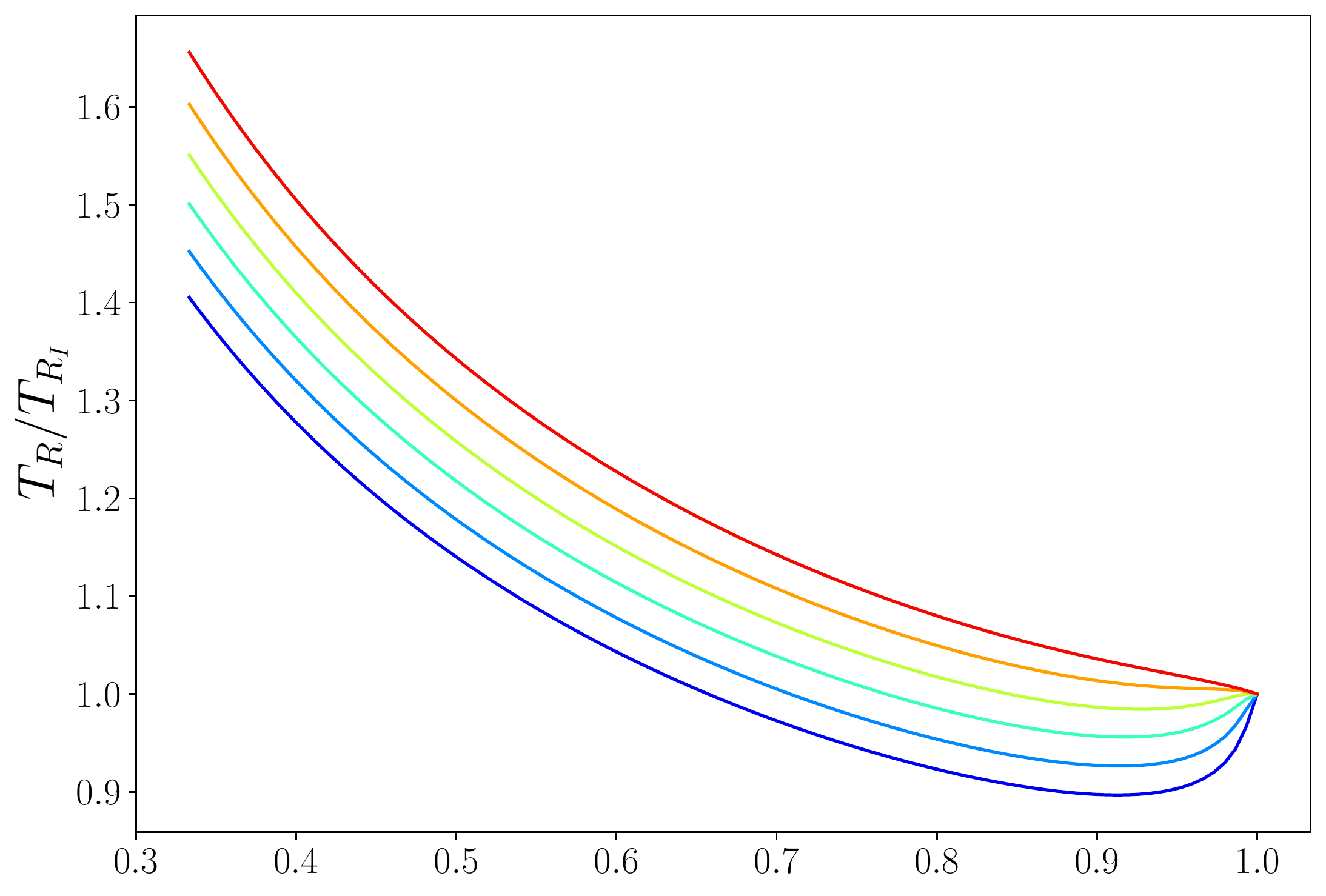} 
    \includegraphics[width=\linewidth]{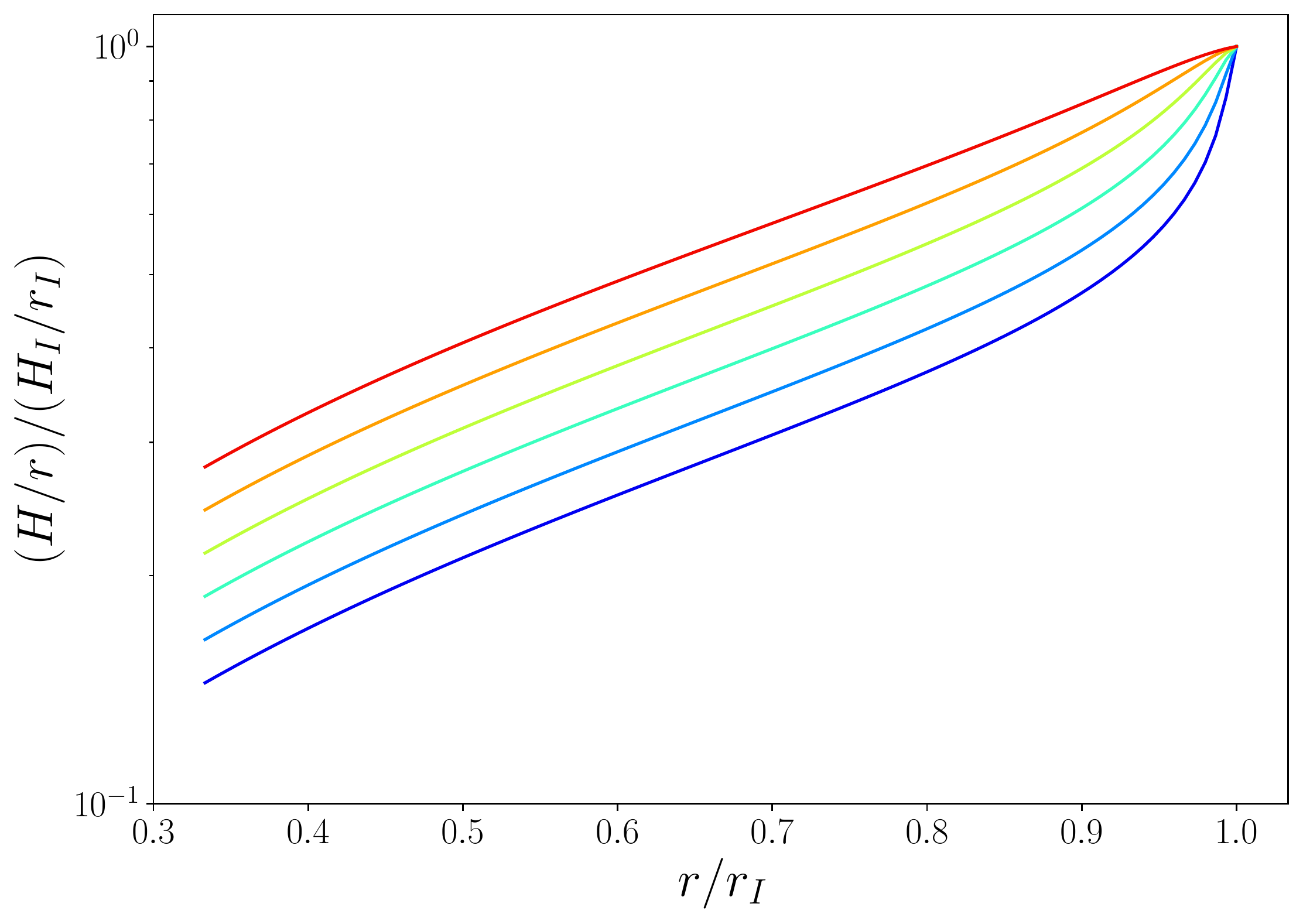} 
     \caption{ Same as Fig. \ref{fig1}, except with $\Gamma = 4/3$ (i.e., a radiation  pressure dominated flow).   } 
 \label{fig2}
\end{figure}

In Figures \ref{fig1} and \ref{fig2} we plot, for a number of different trans-ISCO velocities $u_I$, the properties of the total pressure $P$, the surface density $\Sigma$, the radiative temperature $T_R$ and the disc aspect ratio $H/r$, normalised by their ISCO values, throughout the intra-ISCO region.  Note the linear vertical scale on the $T_R$ plot, in contrast with the logarithmic vertical axis of the other plots. In Figure \ref{fig1} we assume that the gas pressure dominates throughout the plunging region ($\Gamma = 5/3$), while in Figure \ref{fig2} we assume that the radiation pressure dominates ($\Gamma = 4/3$). In both Figures we assume that the electron scattering opacity dominates throughout. The horizontal axis of both Figures spans $1$ to $1/3$, i.e., the range which would be relevant for a Schwarzschild black hole. 

There are a number of interesting properties of these solutions. First, note the strong dependence on the properties of the solution on the trans-ISCO velocity $u_I$, with larger trans-ISCO velocities resulting in a hotter disc throughout the plunging region.  The solutions with lower trans-ISCO velocities rapidly cool in the near-ISCO region $r \sim 0.95-1 r_I$, a result of the more pronounced radial stretching of the disc fluid, owing to its larger radial acceleration.  Each solution displays a pronounced drop in total pressure across the intra-ISCO region, which is relatively independent of the adiabatic index $\Gamma$. However, mass conservation results in a similarly pronounced drop in $\Sigma$, and therefore also in the optical depth. These competing effects result in a much shallower decay in the radiative temperature $T_R$, which may even rise throughout the intra-ISCO region, despite there being no turbulent heating of the flow in this regime.  It is also interesting to note that the ``thin disc'' assumption is a good one throughout the plunging region, with the disc aspect ratio $H/r$ decreasing with decreasing $r$, a result of the increasing vertical gravity of the near-horizon Kerr geometry. Finally, note that the dominant source of pressure (radiation versus gas) in the disc only slightly affects the evolution of many of the disc thermodynamic quantities (e.g., contrast the total pressure profiles in Figs. \ref{fig1} and \ref{fig2}), but the properties of the radiative temperature are much more sensitively dependent on the choice of adiabatic index $\Gamma$.   Recall that we have neglected radiative losses in producing the Figs. \ref{fig1} and \ref{fig2}, and so these profiles should be considered formal solutions, which will likely be modified by the inclusion of radiative losses.

\section{Discussion}
\subsection{The vertical compression and radial expansion of the intra-ISCO flow}
\begin{figure}
\centering
  \includegraphics[width=\linewidth]{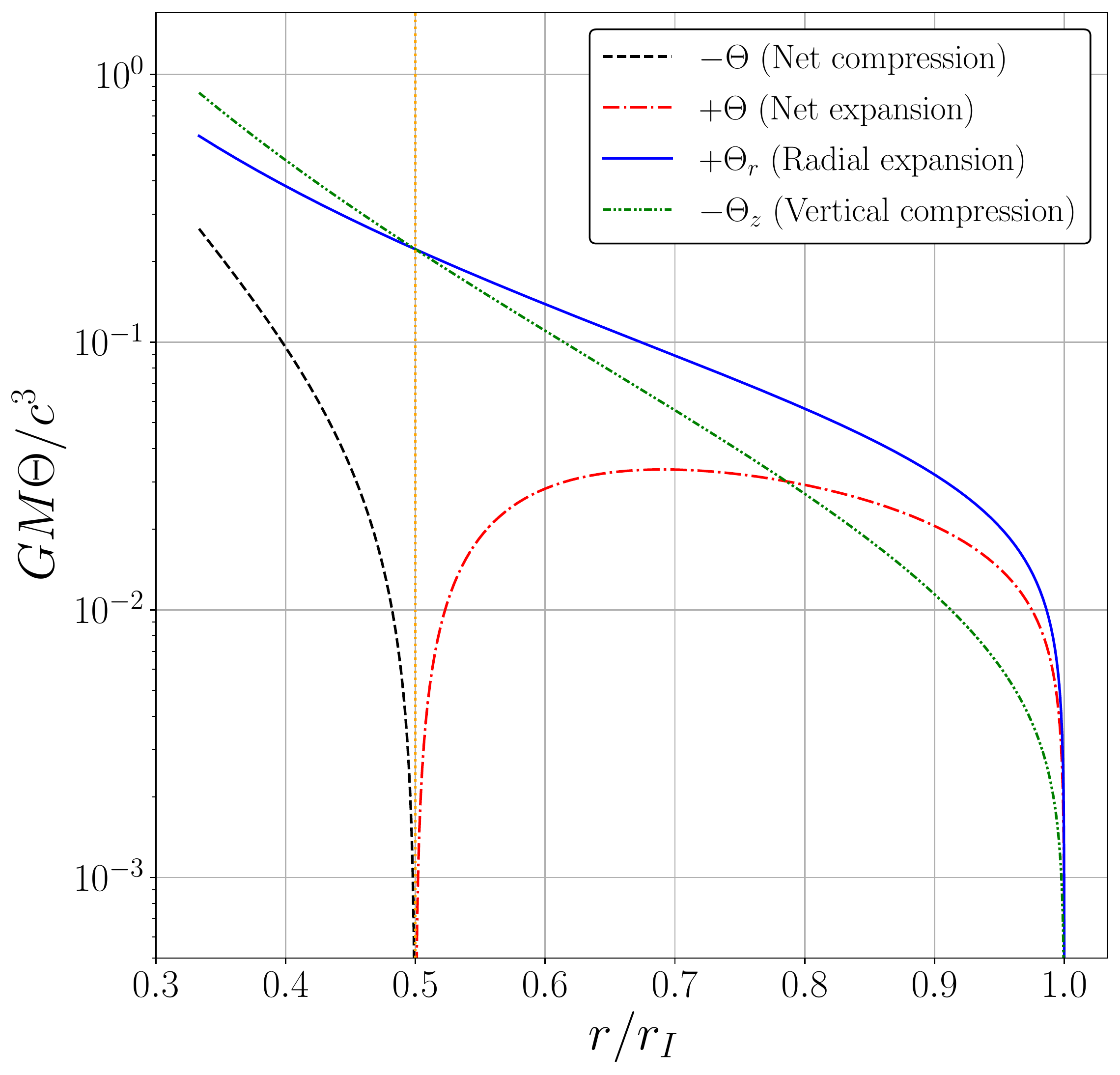} 
   \caption{The components of, and the total value of, the disc compression scalar $\Theta = {\rm d}\ln{\cal V} / {\rm d}\tau$. Figure produced with $a_\star = 0$, $\Gamma = 5/3$.    } 
 \label{fig3}
\end{figure}

As can be seen from Figs. \ref{fig1} \& \ref{fig2}, the pressure (as well as the central temperature and density)  of the intra-ISCO disc solutions initially decreases as it crosses the ISCO, before plateauing and then increasing towards the event horizon. As the evolution of the disc is adiabatic, this implies that the fluid is initially stretched, before being compressed closer to the black hole. The evolution of a volume element can be examined starting with the equation of mass conservation 
\beq
\nabla_{\mu} \left(\rho U^\mu \right) = 0,
\eeq
which expands to give 
\beq
\nabla_{\mu} \left(\rho U^\mu \right) = \rho \nabla_\mu U^\mu + U^\mu \nabla_\mu \rho = \rho \nabla_\mu U^\mu + U^r {\partial \rho \over \partial r} = 0.
\eeq
Rearranging, we find 
\beq\label{comp_def}
\nabla_\mu U^\mu = -{U^r \over \rho } {\partial \rho \over \partial r} = - {{\rm d} \ln \rho \over {\rm d} \tau} = {{\rm d} \ln {\cal V} \over {\rm d} \tau} ,
\eeq
where $\rho \equiv m / {\cal V}$, where $m$ is a (constant) rest mass, and ${\cal V}(r)$ is an elemental volume.    We may examine its properties by computing the discs compression scalar 
\begin{multline}
\Theta \equiv \nabla_\mu U^\mu = {1 \over \sqrt{|g|}} \partial_\mu (\sqrt{|g|} U^\mu ) \\ = {1\over r} {\partial \over \partial r} (r U^r) + {\partial \over \partial z} U^z \equiv \Theta_r + \Theta_z .
\end{multline}
By this definition, $\Theta > 0$ corresponds to expansion and $\Theta < 0$ contraction. The disc compression scalar is given by the sum of two terms, both of which have a clear physical interpretation. The change in disc volume associated with radial motion is simple to write: 
\beq
\Theta_r = {1\over r} {\partial \over \partial r} (r U^r) = {U^r \over r} + {\partial U^r \over \partial r} ,
\eeq
while the vertical compression can be determined from 
\beq
\Theta_z = - {1\over r} {\partial \over \partial r} \left(r U^r \right) - U^r {\partial \ln \rho \over \partial r} ,
\eeq
meaning (using equation \ref{rhrh} to compute the density derivative)
\beq
\Theta_z =  \left({5 - \Gamma\over \Gamma + 1}\right) {U^r \over r} - \left({\Gamma - 1 \over \Gamma + 1}\right) {\partial U^r \over \partial r} .
\eeq
It is not surprising to note that the radial expansion of the fluid is $\Gamma$-independent, as it is driven entirely by gravity. The vertical compression is $\Gamma$-dependent however, as it results from the balance of vertical gravity and the disc pressure. The total compression scalar is given by  
\beq
\Theta = \Theta_r + \Theta_z = {6 \over \Gamma + 1} {U^r \over r} + {2 \over \Gamma + 1} {\partial U^r \over \partial r} .
\eeq
Within the ISCO we have {(with corrections of order $u_I/c \ll 1$)}
\begin{align}
U^r &\simeq - c \sqrt{2r_g \over 3 r_I} \left({r_I \over r} - 1\right)^{3/2}, \\ {\partial U^r \over \partial r} &\simeq  {3 c \over 2r_I} \sqrt{2r_g \over 3 r_I} \left({r_I \over r}\right)\left({r_I \over r} - 1\right)^{1/2} ,
\end{align}
meaning that both the radial- and vertical- compression are simple polynomials in $r_I/r$. These functions are plotted in Figure \ref{fig3}. In Figure \ref{fig3} we see that the radial expansion (blue solid curve) initially dominates over vertical compression (green curve), leading to a net radial expansion (red dot-dashed curve). However, closer to the event horizon the vertical compression wins out over the radial expansion, leading to a net compression of the fluid (black dashed curve). The crucial point at which vertical compression overcomes radial expansion is when $\Theta = 0$, or 
\beq
{\partial \ln U^r \over \partial \ln r} \simeq -3 \rightarrow - \left({r_I \over r} - 1\right) +{1\over 2} {r_I \over  r} \simeq 0 \rightarrow r \simeq {r_I \over 2} .
\eeq
Thus for discs around black holes with spins $a_\star < 0.687\dots$ the fluid is compressively heated before reaching the event horizon  {(this spin corresponds to the value at which the outer event horizon precisely equals $r_I/2$)}. 
A final note is that we have neglected the small trans-ISCO velocity $u_I$ in computing the solution $\Theta = 0$, if we were to include it we find the first order correction to this result is 
\beq
r = r_I \left({1\over 2} + \varepsilon\right) + {\cal O}(\varepsilon^2) .
\eeq	
The zero net compression radius moves slightly towards the ISCO for a non-zero $u_I$.  {It is interesting to note that an increase in the temperature within $r < r_I/2$ was also found in the numerical simulations of Schnittman et al. (2016). More detailed comparisons are required before identifying whether this increase results from the same physical mechanism as spelled out in this paper however. }

\subsection{The trans-ISCO velocity} 
As is clear to see from Figures \ref{fig1} and \ref{fig2}, the trans-ISCO velocity $u_I$ represents a crucial physical parameter for determining the physical and observed  properties of the intra-ISCO flow. Classical relativistic accretion models (e.g., the Novikov, Page \& Thorne (1973, 1974) solutions) allow the mean second-order radial accretion velocity of the flow at a given radius to be determined as a function of the physical parameters of the system (e.g., the black hole mass and spin, the mass accretion rate, and the disc stress $\alpha$-parameter). This at first appears to be a likely candidate, once evaluated at the ISCO,  for the trans-ISCO velocity.  

However, it is important to recall that accretion flows are turbulent, and that the typical scale of the turbulent fluctuations exceeds the mean drift velocity by an asymptotic scale 
\beq
|U^r| \ll \delta U^r . 
\eeq
In the main bulk of the disc, these velocity fluctuations vanish on average 
\beq
\left\langle \delta U^r(r\gg r_I) \right\rangle = {1 \over \Delta t} \int_t^{t+\Delta t} \delta U^r(r\gg r_I, t') \, {\rm d}t' = 0. 
\eeq
In effect this statement implies that turbulent fluctuations outwards in the disc are as likely as turbulent fluctuations inwards. 
However, close to the ISCO itself, these velocity fluctuations will develop a non-zero directional bias, with fluctuations across the ISCO in effect absorbed into the black hole due to the large radial gravitational acceleration of the intra-ISCO region. This means that fluctuations across  the ISCO from the main body of the disc are no longer compensated by fluctuations back from the intra-ISCO region. This favouring of fluctuations in a specific direction will mean that the average defined above will no longer vanish in the near-ISCO region, and instead
\begin{multline}
\left\langle \delta U^r(r\sim r_I) \right\rangle = {1 \over \Delta t} \int_t^{t+\Delta t} \delta U^r(r\sim r_I, t') \, {\rm d}t'  \\ \sim - \left| \delta U^r(r \sim r_I) \right| \gg |U^r| .
\end{multline}
Thus, careful attention must be paid to the precise value of the the trans-ISCO velocity used in computing the evolution  of the intra-ISCO thermodynamic quantities.  This will be particularly relevant in the high ISCO stress regime, where the mean flow velocity decreases with ISCO stress $W_I$ (proof in Appendix \ref{AA})
\beq
u_I \propto W_I^{-3/2} .
\eeq
The speed of sound, however, will increase with an increased ISCO stress, a result of the higher inner disc temperatures. The speed of sound is likely a good measure of the typical turbulent velocity fluctuation scale, and therefore while the typical mean drift velocity will decrease, the typical trans-ISCO velocity may well increase as a function of ISCO stress. This conceptual point will be analysed more concretely with a dedicated calculation in a work to follow. 

\subsection{A full solution: joining onto an extra-ISCO accretion flow}
We now have a simple, physically motivated, analytical solution of the intra-ISCO thermodynamic equations.  An obvious first application of this solution is to join it onto an extra-ISCO relativistic disc, therefore creating a global accretion solution valid smoothly from the event horizon out to large radius, a theoretical result of some interest. In this sub-section we construct such a solution. 

As we have mentioned throughout this paper, there are various caveats and assumptions made while constructing our solution which must be kept in mind, and tested for validity. By joining the intra-ISCO flow onto an extra-ISCO disc, these assumptions can be more robustly tested, as the various scales of important parameters (e.g., the relative magnitudes of the accelerations due to gravity and pressure gradients) can be explicitly related to the physical parameters of the model. 

For our outer disc model we take the relativistic solutions of Novikov, Page \& Thorne (1973, 1974).  The global disc solutions take as inputs 5 free parameters: the black hole mass $M$ and spin $a$, the (constant) mass accretion rate through the disc $\dot M$, the disc stress $\alpha$-parameter and an ISCO stress parameter $\delta_{\cal J}$. Recall that, for an accretion flow to have a non-infinite trans-ISCO velocity, there must be a non-zero ISCO stress present. 

We parameterise the ISCO stress by a dimensionless  angular momentum flux  $\delta_{\cal J}$, which physically corresponds to the angular momentum carried back to the stable flow from the intra-ISCO material, in units of the ISCO circular orbit angular momentum (we discuss at length our extra-ISCO disc solutions in Appendix \ref{AA}). We parameterise the ISCO stress in this manner as $\delta_{\cal J}$ is a number readily extractable from simulations, and is typically found to have magnitude  $\delta_{\cal J} \sim 0.01 - 0.1$ (e.g., Noble, Krolik \& Hawley 2010, Schaffe et al. 2008, Penna et al. 2010,  Zhu et al. 2012, Schnittman, Krolik \& Noble 2016).  

\begin{figure*}
    \includegraphics[width=.95\linewidth, height=1.25\linewidth]{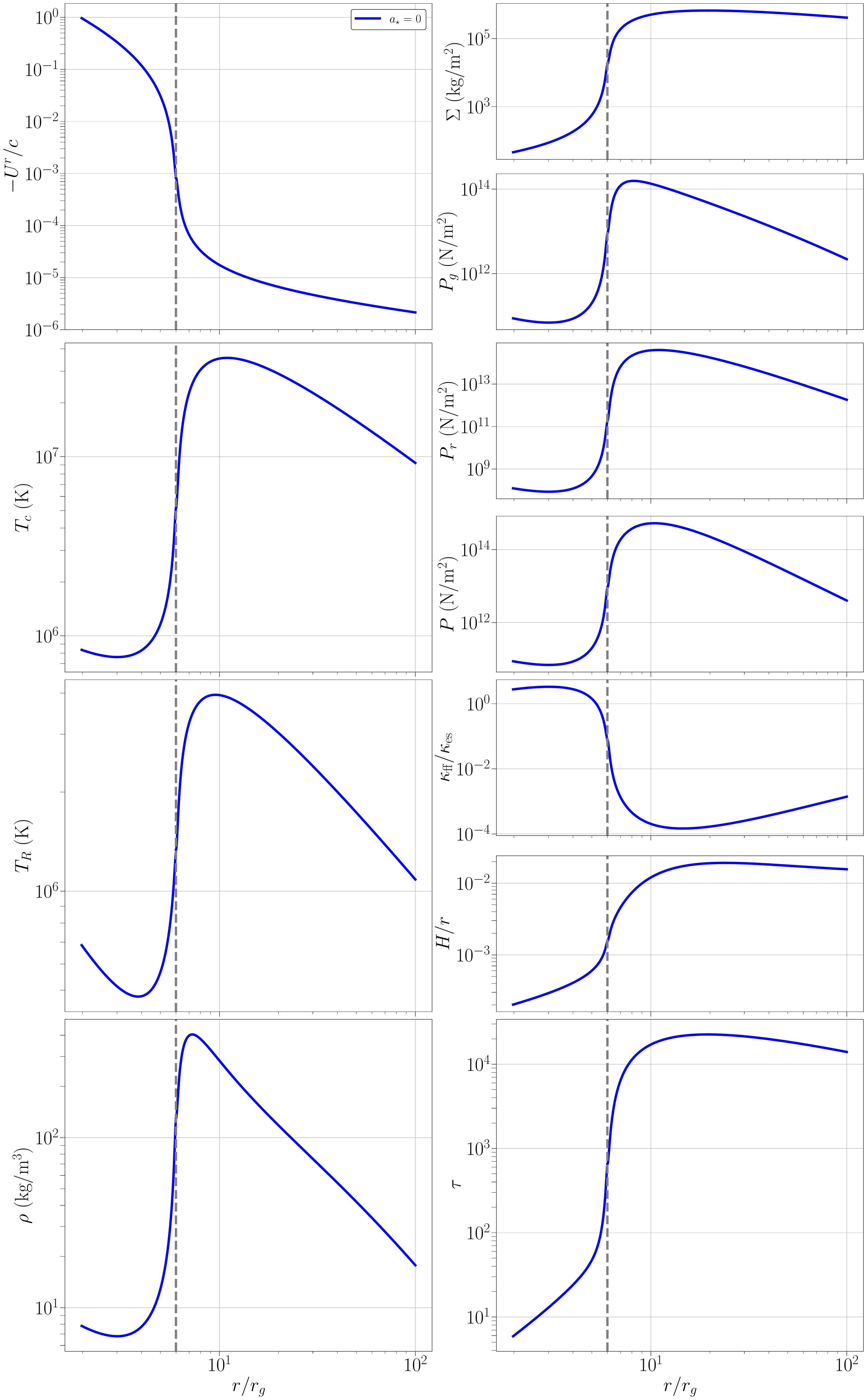} 
     \caption{ The global solution of the disc dynamic and thermodynamic quantities, both outside and within the ISCO (denoted by the grey dashed curve). This Figure was produced for a black hole mass $M = 10 M_\odot$, black hole spin $a = 0$, constant mass accretion rate $\dot M = 0.1 \dot M_{\rm edd}$, alpha parameter $\alpha = 0.1$, and ISCO-stress parameter $\delta_{\cal J} = 10^{-4}$. See text for further discussion.   } 
 \label{fig4}
\end{figure*}

Before we present numerical solutions of our governing equations, we remind the reader here of the approximations and simplifications that are inherent to the extra-ISCO solution (Appendix \ref{AA}). In the outer diffusive disc we have assumed (i) that the electron scattering opacity dominates the total opacity $\kappa_{\rm es} \gg \kappa_{\rm ff}$, and (ii) that the turbulent stress $W^r_\phi$ is proportional to the gas pressure of the disc (i.e., the Shakura-Sunyaev (1973) $\alpha$-formulation which prevents the onset of Lightman-Eardley (1974) instabilities).  Within the ISCO we include both the electron scattering and free-free opacities, which are often comparable.  In both the extra- and intra-ISCO solutions we include the contributions of gas and radiation pressure in the solution to the constraints of hydrostatic equilibrium. The effect of radiation pressure within the ISCO has been discussed  in the proceeding section, but it also contributes meaningfully in the main diffusive regions of the disc. In the extra-ISCO region the equation of hydrostatic equilibrium is
\beq
H = \sqrt{\left({k_B T_c \over \mu m_p} + {4 \sigma T_c^4 H \over 3 \Sigma c} \right) \left({r^4 \over U_\phi^2 + a^2 c^2 (1 - U_0^2)}\right)} ,
\eeq
with $T_c$, $\Sigma$, $U_\phi$ and $U_0$ known functions of radius (once the input parameters have been specified). Thus we have a quadratic to solve, with physical solution 
\beq
H = {r^4 \over L^2 }\left({2\sigma T_c^4 \over 3 \Sigma c}\right) + \sqrt{{r^8 \over L^4} \left({2\sigma T_c^4 \over 3 \Sigma c}\right)^2 + \left({k_B T_c \over \mu m_p}\right){r^4 \over L^2} }  ,
\eeq
where we have defined the short hand $L^2 \equiv U_\phi^2 + a^2 c^2 (1 - U_0^2)$. With each thermodynamic and dynamic quantity specified by the free parameters of the model, the ISCO values of these quantities are then used as boundary conditions for the intra-ISCO flow.

Within the ISCO we numerically solve the adiabatic evolution equation of the intra-ISCO flow, namely 
\beq
{{\rm d}e \over {\rm d} r} - (P + e) {{\rm d} \ln \rho \over {\rm d} r} = 0, 
\eeq
where 
\beq
e = {3 \over 2} P_g + 3 P_r, \quad P = P_g + P_r, 
\eeq
and the density $\rho$ is given by $\Sigma/H$, with $\Sigma$ and $H$ described by the equations given in the previous section. This equation is trivial to solve numerically, and has the benefit of including the contributions of both the radiation and gas pressures simultaneously in the intra-ISCO region. Note that for all of the figures shown in the rest of this paper, the numerical solutions deviate from the analytical solutions of the previous section by at most $1\%$. 

In this section, we do not compute solutions for all of the  parameter space formally available to the models. This is because we have argued, for example, that the trans-ISCO velocity will be severely underestimated by classical models in the high ISCO stress regime. Therefore, for now, we focus on solutions with low values for the ISCO stress, where the trans-ISCO velocity is of order the speed of sound, as expected from general physical arguments. In Figure \ref{fig4} we plot our ``canonical'' low-stress model, the accretion solution for a 10 solar mass Schwarzschild black hole, with accretion rate $\dot M = 0.1 \dot M_{\rm edd}$ and $\alpha = 0.1$. We take an ISCO stress parameter $\delta_{\cal J} = 10^{-4}$. The ISCO radius $r_I = 6GM/c^2$ is denoted by the vertical dashed grey line. 



In Figure \ref{fig4} we plot all relevant dynamic and thermodynamic quantities of the accretion flow from an outer radius of $r = 100 GM/c^2$, down to the event horizon (although formally the event horizon is not a special location in the thermodynamic equations, and does not represent a singularity of the thermodynamics).  It is remarkable how smoothly the solution within the ISCO and outside the ISCO appear to join. With this global solution constructed, we can test a number of the simplifying assumptions employed in deriving these solutions. 

Firstly, it is clear that while the optical depth of the intra-ISCO material drops sharply, the flow remains formally optically thick with $\tau > 1$. The rapidly decreasing surface density $\Sigma$ (a result of mass conservation and a rapid radial expansion) is somewhat compensated by a growing free-free opacity $\kappa_{\rm ff}$ throughout the plunging region, the $\tau \propto r^2$ relationship in a free-free dominated regime is recovered close to the event horizon. This means that the optically thick relationship between radiative and central temperatures (eq. \ref{opt_thick}) is a robust one. 

\begin{figure}
\centering
  \includegraphics[width=\linewidth]{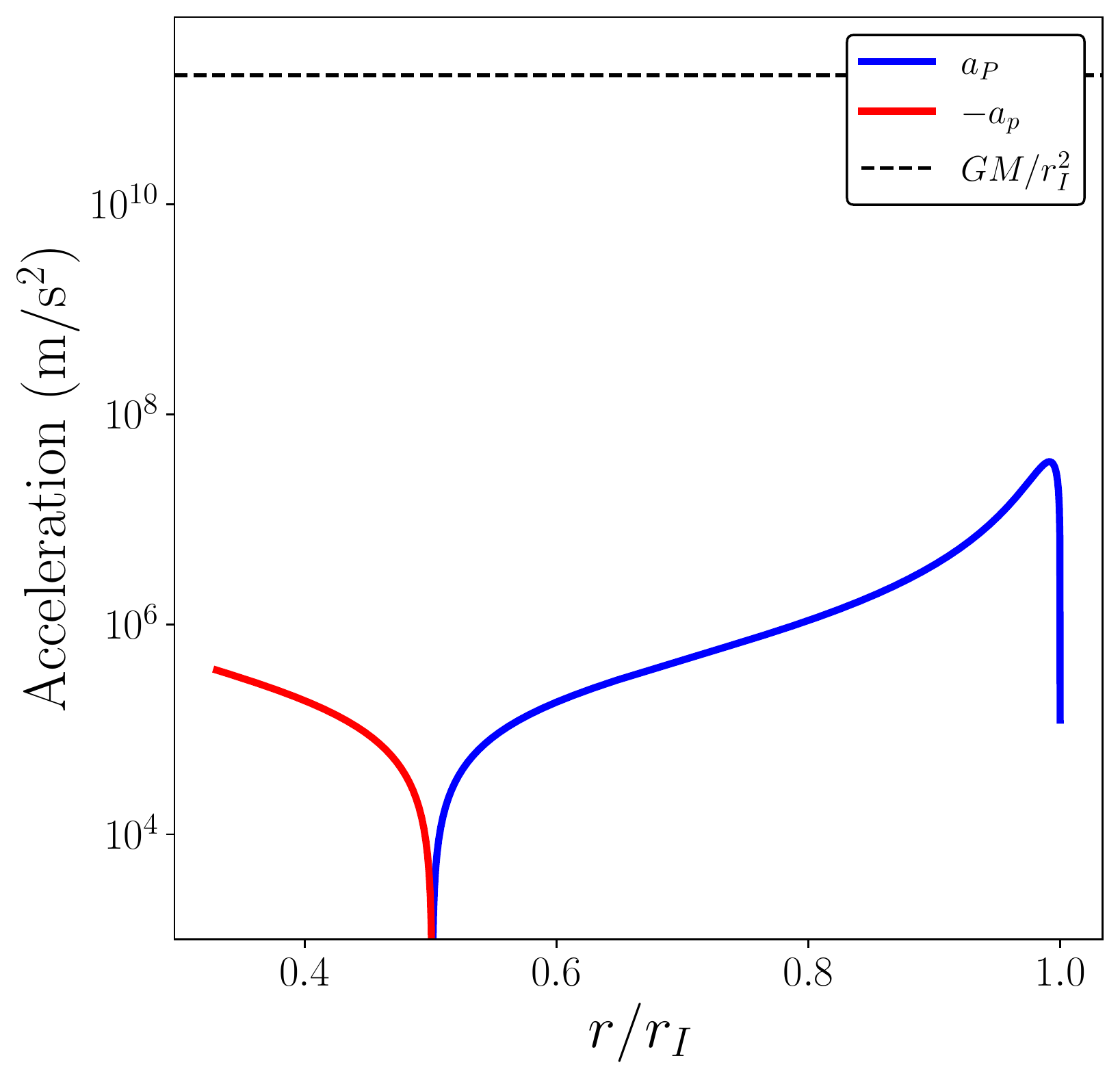} 
   \caption{ The scales of the two relevant acceleration terms in the relativistic Euler equation. The acceleration due to pressure gradients, defined by $a_P = {1/\rho} \, {\rm d}P/{\rm d}r$ is many orders of magnitude smaller than the gravitational acceleration scale $GM/r_I^2$. The quantity $a_P$ was computed numerically from the Schwarzschild  intra-ISCO solutions of Fig. \ref{fig3}. This confirms that the geodesic solution of the relativistic Euler equations (eq. \ref{flow_soln}) is an extremely accurate one.   } 
 \label{fig5}
\end{figure}

Secondly, the thin disc assumptions are also robust, with $H/r \ll 1$ satisfied throughout the intra-ISCO region. Indeed, the disc aspect ratio decreases monotonically with decreasing radius. Similarly, the geodesic  orbital solutions of the relativistic Euler equation is a safe assumption, as the   ``acceleration due to pressure gradients'' is significantly smaller than the typical gravitational acceleration scale (see Fig. \ref{fig5}). 

\begin{figure}
\centering
  \includegraphics[width=\linewidth]{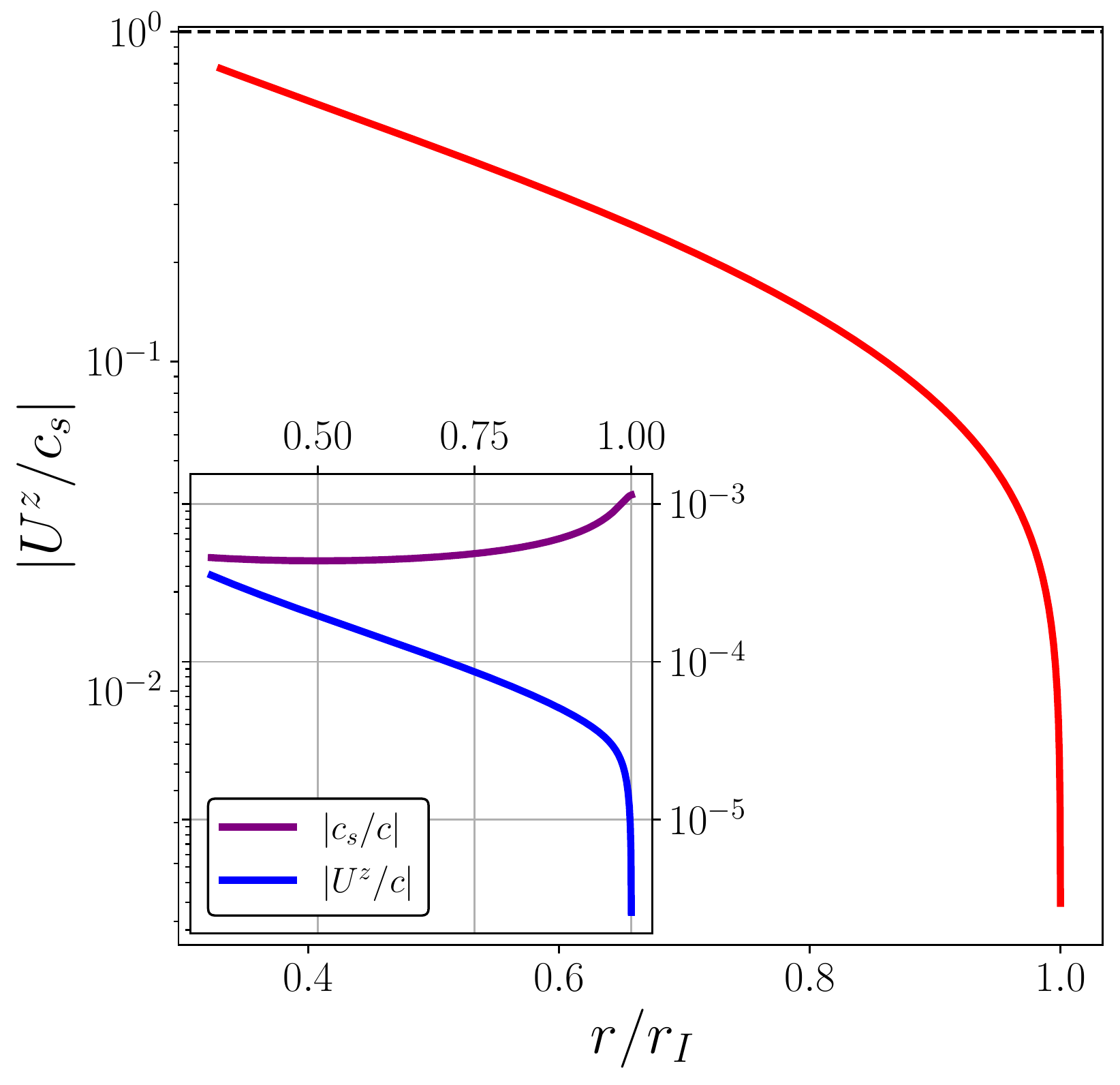} 
   \caption{ The ratio of the vertical velocity and the speed of sound of the trans-ISCO flow, as a function of normalised radius. Inset: the speed of sound and vertical velocity as a fraction of the speed of light.  Both  quantities were computed numerically from the Schwarzschild  intra-ISCO solutions of Fig. \ref{fig3}. The vertical velocity required to maintain vertical hydrostatic equilibrium is sub-sonic in the intra-ISCO region, meaning that the flow will remain in vertical hydrostatic equilibrium  } 
 \label{fig6}
\end{figure}

Finally, the assumption of strict vertical hydrostatic equilibrium is also robust, for this black hole spin. In Figure \ref{fig6}, we plot the ratio  $|U^z| / c_s$ of the inner flow, computed numerically from the solutions displayed in Figure \ref{fig4}. The speed of sound is defined by
\beq
c_s \equiv \sqrt{{\rm d} P \over {\rm d}\rho} ,
\eeq
while the vertical velocity required to maintain vertical hydrostatic equilibrium is given by 
\beq
U^z \equiv {{\rm d} H \over {\rm d}\tau} = U^r {{\rm d} H \over {\rm d} r} . 
\eeq
In Fig. \ref{fig6} we find that the vertical velocity required to maintain vertical hydrostatic equilibrium is sub-sonic for all radii in the intra-ISCO region which lie  outside of the event horizon.  In the following section we shall demonstrate that this is not necessarily the case for all black hole spins, and in particular large retrograde spins will not satisfy this assumption at all radii. This of course does not mean that the flow for these solutions will undergo shocks in the vertical direction, but instead  that the flow will no longer be able to maintain vertical hydrostatic equilibrium, a process which is mediated by motions travelling at the speed of sound.  We shall re-solve the thermodynamic equations within this ``vertical sonic radius'', assuming that the fluid free-streams in a later section.

There is naturally a vast parameter space which may be explored in these solutions. Understanding the parameter (e.g., black hole mass and spin) dependence of each thermodynamic quantity in the intra-ISCO region is an important question. To prevent cluttering the main body of this paper we present a representative set of example Figures in Appendix \ref{parameter_plots_C}. In these Figures we demonstrate the black hole mass dependence (section E1), black hole spin dependence (section E2), accretion parameter ($\dot M, \alpha$) dependence (section E3) and ISCO stress dependence (section E4), of these solutions. One result of particular interest is highlighted in Appendix E4. Namely, that for high ISCO stresses  the radiative temperature of the flow can globally peak within the ISCO. The values of the ISCO stress required for this intra-ISCO peaking are comparable to those found in GRMHD simulations of thin discs $\delta_{\cal J} \sim 0.01 - 0.1$ (e.g., Noble, Krolik \& Hawley 2010, Schaffe et al. 2008, Penna et al. 2010,  Zhu et al. 2012, Schnittman, Krolik \& Noble 2016). 

{Finally, it is interesting to compare the global solutions presented in this section with the pioneering numerical calculations of P21.   Our approach is different in several ways. The first is on a technical level:  by analytically solving the geodesic inspiral motion we have been able to construct far simpler governing thermodynamic equations compared with those of P21.    As we have demonstrated in the preceding section, the analytic nature of the solution (with accuracies of about a percent), allows for dramatic decrease in the time required for computing solutions in, for example, the fitting of observed X-ray spectra.  Another important feature of this simplification is that it allows us to include  the effects of radiation and gas pressures, and both free-free and electron scattering opacities associated with the intra-ISCO flow.   By contrast, P21 limits itself to gas pressure and electron scattering opacities.   We have seen, however, that neglecting either the radiation pressure or free-free opacities can lead to significant thermodynamic changes (which are observationally relevant) in the calculated intra-ISCO profiles.   This is particularly true for the effects of the radiation pressure.   The current study also differs from P21 in its approach to the calculated stress.    P21 computes the stress using an $\alpha$-model for the disc turbulence within the ISCO. While the modelling of a turbulent stress via an effective viscosity is well-motivated in the main body of the disc, where it is absolutely essential for the accretion process, its role in the intra-ISCO region is far less clear.   And, while there seems little doubt that magnetic stresses will be present at some level in this region (e.g., Noble, Krolik \& Hawley 2010, Schaffe et al. 2008, Penna et al. 2010,  Zhu et al. 2012, Schnittman, Krolik \& Noble 2016), their affect on the post-ISCO relativistic inflow, which need not lose any angular momentum to accrete, is unlikely to be well-captured by a statistically averaged $\alpha$-model.     We have therefore chosen to focus on geodesic flows here, leaving the question of dynamical stresses to be more fully addressed by detailed comparisons with dedicated numerical simulations.  

Lastly, the assumption of strict vertical hydrostatic equilibrium must be revisited for some regions of interest in parameter space, as we now will show.  }

\section{Intra-ISCO thermodynamics in the free-streaming regime }
As we demonstrated in the proceeding section, the global accretion solutions constructed in this paper require, for reasonable values of the free parameters of the theory, a vertical velocity which becomes of order the sound velocity in the very innermost regions of the black holes spacetime (see e.g., Fig \ref{fig6}). We reiterate that a trans-sonic vertical velocity would not imply that the intra-ISCO fluid will undergo vertical shocks, but simply that the fluid will no longer be able to maintain vertical hydrostatic equilibrium. Inside of this region the fluid will instead ``free-stream'' and undergo a spherical plunge $H(r<r_s) = H(r_s) (r/r_s)$.  

We now demonstrate that this free-streaming radius is, for a given adiabatic parameter $\Gamma$, a constant fraction of the ISCO radius $r_I$, independent of all other parameters. This means that only for certain black hole spins will this location occur outside of the black hole's event horizon.  

\subsection{The location of the free-streaming radius} 
\begin{figure}
\centering
  \includegraphics[width=\linewidth]{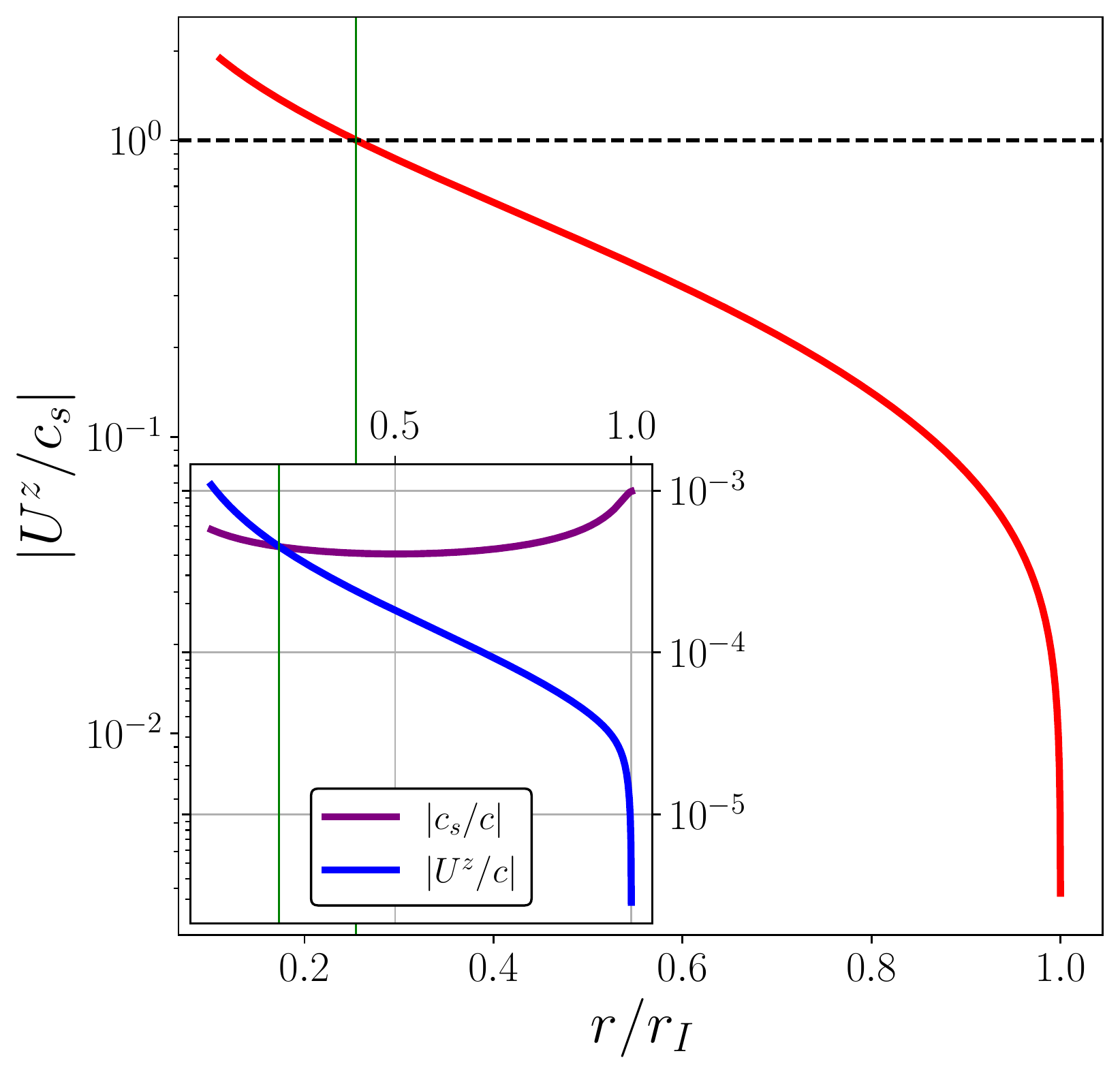} 
   \caption{ The ratio of the vertical velocity and the speed of sound of the trans-ISCO flow, as a function of normalised radius. Inset: the speed of sound and vertical velocity as a fraction of the speed of light. The green vertical curves in both plots are located at $r = 0.25458r_I$.  Both  quantities were computed numerically from the  intra-ISCO solutions described in the previous section, with $a = -M$. The vertical velocity required to maintain vertical hydrostatic equilibrium becomes super-sonic in the innermost intra-ISCO region, meaning that the flow will deviate from vertical hydrostatic equilibrium.  } 
 \label{fig8}
\end{figure}

We can determine the location of the free-streaming radius from the expression derived in section \ref{iscobondi}, namely:
\beq
{U^z \over c_s } = {1\over \sqrt{\Gamma}} {U^r \over c} \left[ 2 +  {\partial \ln c_s \over \partial \ln r}  \right] \sqrt{r^2 \over 2 r_g r_I} .
\eeq
The free-streaming radius $r_s$ is then given by the  solution of 
\begin{multline}
\left| \sqrt{1 \over \Gamma} {U^r(r_s) \over c} \sqrt{r_s^2 \over 2 r_g r_I} \left(2 + \left.{\partial \ln c_s \over \partial \ln r}\right|_{r_s}\right)\right|  \\ = {1\over \sqrt{\Gamma}} \sqrt{2r_g \over 3 r_I} \left( {r_I \over r_s} - 1\right)^{3/2}\sqrt{r_s^2 \over 2 r_g r_I} \left(2 + \left.{\partial \ln c_s \over \partial \ln r}\right|_{r_s}\right)  = 1.
\end{multline}
Note that the $r_g$ dependence of this expression drops out, and we are left with 
\beq
\sqrt{1 \over 3 \Gamma} \left({r_s \over r_I} \right)\left( {r_I \over r_s} - 1\right)^{3/2} \left(2 + \left.{\partial \ln c_s \over \partial \ln r}\right|_{r_s}\right)  = 1.
\eeq
As with all quantities in the intra-ISCO regime, the adiabatic solution for the speed of sound is simply a polynomial in $r/r_I$ throughout the region $r < r_I$, given explicitly by 
\beq
c_s = \sqrt{\Gamma P \over  \rho} = c_{s, I}  \left({r_I \over r} \right)^{{3\Gamma-3 \over \Gamma + 1}} \left[ \varepsilon^{-1} \left({r_I \over r} - 1\right)^{3/2} + 1\right]^{-{(\Gamma-1)\over (\Gamma + 1)}}.
\eeq
As $\varepsilon \ll 1$, at the free-streaming radius we are able to neglect the term of order $1$ compared to the term of order $1/\varepsilon$, and thus the speed of sound gradient $\partial \ln c_s /\partial \ln r$ will depend only on $r_s/r_I$, and not on the trans-ISCO velocity.  This means that the free-streaming radius is a function of only $\Gamma$ and $r_s/r_I$, and for a given adiabatic  law will occur at a fixed fraction of the ISCO radius, independent of black hole spin. For $\Gamma = 5/3$ this radius corresponds to
\beq
{r_s \over r_I} \simeq 0.25458\dots 
\eeq
This means that for black hole spin parameters 
\beq
a_\star < -0.456\dots
\eeq
this location is outside of the black holes event horizon. This is verified in Figure \ref{fig8}, where we plot the vertical and sound velocities  of the intra-ISCO flow, computed numerically from a global disc model with $a = -M$. The other physical parameters where kept the same as Fig. \ref{fig4}. 

 \begin{figure*}
    \includegraphics[width=.75\linewidth, height=1\linewidth]{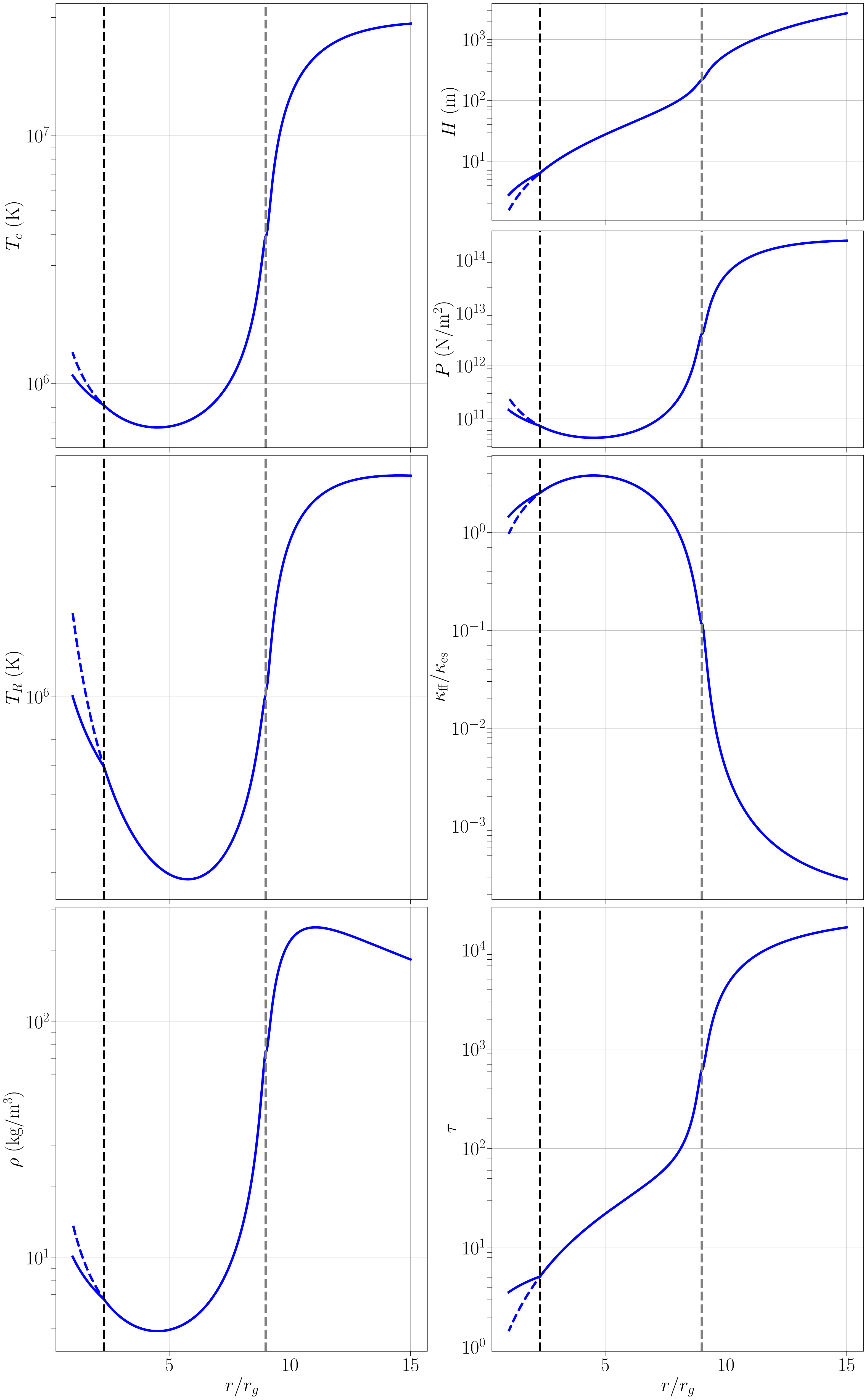} 
     \caption{ The global solution of the disc dynamic and thermodynamic quantities, both outside and within the ISCO (denoted by the grey dashed curve), now with the inclusion of the free-streaming zone. The free-streaming radius is denoted by the black vertical dashed cure.   This Figure was produced for the same physical parameters as Fig. \ref{fig4}, except now with a black hole spin parameter $a = -M$. The free-streaming region acts to keep the disc cooler than the purely adiabatic solution (displayed by dashed curves). } 
 \label{fig9}
\end{figure*}

\subsection{Behaviour within the free-streaming radius}

If we assume that upon crossing the vertical trans-sonic point the flow thereafter ``free-streams'' and undergoes a spherical plunge (because the information regarding vertical hydrostatic equilibrium  can no longer propagate quickly enough across the thickness of the flow), then the solutions of the thermodynamic equations simplify once more. The density of the flow then simply follows from mass conservation, and within the free-streaming  radius satisfies 
\beq
{\rho(r) \over \rho(r_s) } = {\Sigma(r) r_s \over \Sigma(r_s) r} = \left({r_s^2 U^r(r_s) \over r ^2 U^r(r)}\right) \simeq \sqrt{r_s \over r}\left( {r_I - r_s \over  r_I - r } \right)^{3/2} ,
\eeq
where $r_s$ is the radius of the vertical sonic point, and we have assumed that the radial velocity at the free-streaming point far exceeds the trans-ISCO velocity (a safe assumption).  From this density an adiabatically evolving flow then has pressure given by 
\beq
{P(r) \over P(r_s)} = \left({\rho(r) \over \rho(r_s) }\right)^\Gamma \simeq \left({r_s \over r} \right)^{\Gamma/2}\left( {r_I - r_s \over  r_I - r } \right)^{3\Gamma/2} ,
\eeq
from which the central temperature, radiative temperature, and all other relevant quantities can be trivially computed. Once again, we must specify which of the gas or radiation pressure, and which of the free-free or electron scattering opacities dominates, before constructing exact solutions.

\subsection{Gas pressure supported systems}
 When the gas pressure dominates the flow, we have 
 \beq
 P \simeq P_g = {k_B \rho T_c \over \mu m_p}, \quad e \simeq {3\over 2} P, \quad \Gamma = {5 \over 3} ,
 \eeq
 and therefore a central temperature which falls as 
 \beq
 {T_c \over T_{c, s}} = {P \over P_s } {\rho_s \over \rho} = \left({r_s\over r}\right)^{1/3} \left({r_I - r_s \over r_I - r}\right) ,
 \eeq
 where subscript $s$ denotes the values of various quantities at the free-streaming radius. Assuming that the electron-scattering opacity dominates within the flow we find a radiative temperature which evolves according to 
 \beq
 {T_R \over T_{R, s}} = {T_c \over T_{c, s}} \left({\Sigma_s \over \Sigma}\right)^{1/4} = \left({r_s\over r}\right)^{11/24} \left({r_I - r_s \over r_I - r}\right)^{5/8} .
 \eeq
 We  find that the free-free opacity falls throughout the free-streaming region:
 \beq
 {\kappa_{\rm ff} \over \kappa_{{\rm ff}, s}}  = {\rho \over \rho_s} \left({T_c \over T_{c, s}}\right)^{-7/2} = \left({r_s\over r}\right)^{-2/3} \left({r_I - r_s \over r_I - r}\right)^{-2} ,
 \eeq
 and therefore, in the $\kappa_{\rm ff, s} \gg \kappa_{\rm es}$ regime the optical depth falls as 
 \beq
{ \tau \over \tau_s} = {\kappa_{\rm ff} \Sigma \over \kappa_{{\rm ff}, s} \Sigma_s} =  \left({r_s\over r}\right)^{-7/6} \left({r_I - r_s \over r_I - r}\right)^{-1/2} , 
 \eeq
 with corresponding radiative temperature change of 
 \beq
  {T_R \over T_{R, s}} = {T_c \over T_{c, s}} \left({\tau_s \over \tau}\right)^{1/4} = \left({r_s\over r}\right)^{5/8} \left({r_I - r_s \over r_I - r}\right)^{9/8} .
 \eeq

 \subsection{Radiation pressure supported systems}
 If the radiation pressure dominates over the gas pressure in the disc, then we have the following simplification: 
\beq
P \simeq P_r = {4 \sigma T_c^4 \over 3 c} , \quad e \simeq 3 P_r,  \quad \Gamma = {4 \over 3}.
\eeq
 The central temperature therefore falls as 
 \beq
 {T_c \over T_{c, s}} = \left({ P \over P_s} \right)^{1/4} =  \left({r_s\over r}\right)^{1/6} \left({r_I - r_s \over r_I - r}\right)^{1/2} ,
 \eeq
 while the radiation pressure falls as 
 \beq
  {T_R \over T_{R, s}} = {T_c \over T_{c, s}} \left({\Sigma_s \over \Sigma}\right)^{1/4} = \left({r_s\over r}\right)^{2/3} \left({r_I - r_s \over r_I - r}\right)^{1/8} ,
 \eeq
 which is a shallower fall off than in the gas pressure dominated regime. 
 In the radiation pressure dominated regime  we are safe to assume that $\kappa_{\rm es} \gg \kappa_{\rm ff, s}$, as the free-free opacity slowly falls in this limit 
 \beq
  {\kappa_{\rm ff} \over \kappa_{{\rm ff}, s}}  = {\rho \over \rho_s} \left({T_c \over T_{c, s}}\right)^{-7/2} = \left({r_s\over r}\right)^{-1/12} \left({r_I - r_s \over r_I - r}\right)^{-1/4} . 
 \eeq

 \subsection{Example numerical solution}
 These solutions are plotted in Fig. \ref{fig9}, where we compute a global disc solution for a black hole with mass $M = 10M_\odot$, and spin $a = -M$, accretion rate $\dot M = 0.1 \dot M_{\rm edd}$, alpha-parameter $\alpha = 0.1$, and ISCO stress parameter $\delta_{\cal J} = 10^{-4}$. The Novikov, Page, and Thorne solution  is used from $r = 15 GM/c^2$ down to the ISCO $r_I = 9 GM/c^2$. Within the ISCO the thermodynamics are first described by the adiabatic solutions of the previous section, up until the free-streaming radius $r_s \simeq 2.29 GM/c^2$, whereafter they are described by the free-streaming solution until the event horizon $r_H = GM/c^2$. Note that the free-streaming region acts to keep the disc cooler than the purely adiabatic solution (which is displayed by dashed curves).

\section{Conclusions }
In this paper we have taken the first steps towards an analytical theory describing thin accretion flows in the intra-ISCO region of Kerr black holes. We have focussed our attention on the limit of adiabatic evolution, or equivalently the limit of short free-fall timescales from the ISCO.  This simplifying assumption allows exact, and simple, analytical profiles of the disc properties to be determined, allowing a  clear picture of the physical processes at play in this region to be developed. We have demonstrated how the competing effects of a dropping optical depth and growing radial expansion and vertical compression of the flow inter-compete and lead to a non-trivial evolution of the radiative temperature within the ISCO. 

A result of some importance is the realisation that the intra-ISCO accretion flows are fundamentally self-similar in nature, depending on radius only through the ratio $r/r_I$. This recent has numerical support, and was observed in the numerical simulations of Schnittman et al. (2016). We have also shown, for example, that the central temperature of these solutions decreases for radii $1/2 < r/r_I < 1$, before increasing  for $0 < r/r_I< 1/2$, independent of black hole spin and assumptions regarding the equation of state of the accretion flow. These results highlight how the black hole spin is, even in the intra-ISCO region, a sub-dominant parameter, and effects the solutions only through its scaling of the ISCO radius. 

These new solutions allow global (i.e., valid at all radii) accretion solutions to be constructed, a result of real theoretical and observational interest. The solutions derived in this paper should be simple to implement in numerical X-ray spectral fitting packages, and may offer insight into the relative importance of the intra-ISCO region in understanding the X-ray emission from the population of black hole accretion sources. It seems likely that at least some sources will inhabit regions of parameter space with potentially detectable intra-ISCO emission, and if so these sources will likely represent prime targets for understanding accretion properties in the strong-gravity limit, and may offer strong constraints on parameters such as the ISCO stress and black hole spin. 

During this analysis we have highlighted the importance of a new physical parameter -- the trans-ISCO velocity $u_I$. Standard extra-ISCO theories with a finite ISCO stress do allow this parameter to be determined explicitly, but it is not clear whether these existing models will accurately capture the physical behaviour of this quantity.  For example, increasing the ISCO stress in classical models acts, in the region surrounding the ISCO,  both to decrease the flows radial drift velocity, while increasing the flows speed of sound.   As the radial velocity of an accretion flow is fundamentally turbulent, with fluctuations roughly of order the speed of sound, it is not clear that the typical velocity with which a fluid element crosses the ISCO will be well described by the mean flow. It may well instead be better described by the typical fluctuation scale $c_s$. The important question of determining  the exact behaviour of $u_I$ will be discussed in a dedicated follow up analysis. 

Further improvements to the analysis presented in this paper would involve the inclusion of both radiative losses out of the disc surface, and radiative diffusion from across the ISCO. This analysis would likely be numerical in nature, but would use the governing  intra-ISCO thermodynamic equations derived in section \ref{deriv_therm} of this paper. 

Finally, it is interesting to note that the solutions  constructed in this paper represent a fundamentally new kind of analytical accretion flow solution: those which are both non-radial and non-circular in character. This class of solutions can be thought of as bridging a gap between the classical angular momentum dominated accretion disc flows (e.g., Lynden-Bell \& Pringle 1974, Page \& Thorne 1974), and the angular momentum-less Bondi accretion flow (Bondi 1952). 

\section*{Acknowledgments} 
It is a pleasure to acknowledge stimulating discussions with R. Fender, C. Gammie, J. Krolik, H. Latter,  G. Oglivie, C. Reynolds,  J. Schnittman, and J. Stone.  We thank the reviewer for a careful reading of the manuscript, and a number of useful suggestions which improved the manuscript.  
 The authors are grateful to C. Terquem for  discussions regarding  the free-streaming zone. This work was supported by a Leverhulme Trust International Professorship grant [number LIP-202-014]. For the purpose of Open Access, AM has applied a CC BY public copyright licence to any Author Accepted Manuscript version arising from this submission. This work is partially supported by the Hintze Family Charitable Trust and STFC grant ST/S000488/1.

\section*{Data accessibility statement}
No observational  data was used in producing this manuscript.

\appendix{}

\section{The extra-ISCO thin disc accretion solution}\label{AA}
In this Appendix we formulate the solutions of the extra-ISCO accretion equations in a manner which both simplifies the ISCO joining procedure, and also highlights the physical dependence of important parameters on the ISCO stress. This section uses the Novikov, Page and Thorne and Shakura-Sunyaev formalisms of the accretion flow equations.

In this Appendix we do not present a first-principles derivation of the three governing equations of the flow, for this we direct the reader to Balbus (2017) for a fully time-dependent derivation. 
\subsection{Theoretical preliminaries: definitions and the classical thin disc equations }
The three equations of mass, angular momentum and energy conservation in a relativistic thin disc allow three quantities pertinent to the disc dynamics and energetics to be determined. These are, the energy radiated from the discs upper and lower surfaces:
\begin{equation}\label{energybalaaa}
{\cal F}_{\cal E} = 2\sigma T_R^4=  -   \Sigma W^r_\phi U^0 \,  {{\rm d} \Omega \over {\rm d} r}  ,
\end{equation}
the radial velocity of the accretion flow:
\beq\label{radvelaaa}
 U^r  =  -  {U^0 \over r \Sigma U_\phi '}  {\partial \over \partial r} \left({ r \Sigma W^r_\phi \over U^0} \right) ,
\eeq
and the evolution of the discs surface density: 
\beq
{\partial \Sigma \over \partial t} = {1 \over r U^0} {\partial \over \partial r} \left({U^0 \over U_\phi '} {\partial \over \partial r} \left({r \Sigma W^r_\phi \over U^0} \right) \right) . 
\eeq
Here $\Sigma$ is the disc surface density, $W^r_\phi$ is the $r$-$\phi$ component of the discs turbulent stress tensor, $U^\mu$ is the disc fluids mean 4-velocity, and $U_\mu$ the fluids mean 4-momentum. The rotational frequency of the flow is $\Omega = U^\phi / U^0$, and $U_\phi '$ is the discs angular momentum {\it gradient}.   Note that each of these expressions are fully time dependent and valid out of in-flow equilibrium. 
\subsection{Steady state solutions }
In the steady state, which we now focus on, $\partial_t \Sigma \equiv 0$ and the constraints of mass and angular momentum conservation become
\beq
{\partial \over \partial r} \left( {U^0 \over U_\phi '} {\partial \zeta \over \partial r} \right) = 0 , \quad \zeta \equiv {r \Sigma W^r_\phi \over U^0},
\eeq
where $\zeta$ is a conveniently defined quantity which describes the angular momentum flux through the disc. The solution of this equation is:  
\beq
{\partial \zeta \over \partial r} = C_1 { U_\phi ' \over U^0 }
\eeq
and
\beq
\zeta(r) = C_1 \int^r {U_\phi'  \over U^0 } \, {\rm d}r + C_2 .
\eeq
This integral can be written explicitly in terms of elementary functions (equation \ref{zeta_sol}), but the explicit form is not required for the following general discussion.    This two integration constants $C_1$ and $C_2$ relate to the constant mass $\dot M$ and angular momentum ${\cal F}_{\cal J}$ fluxes through the disc respectively: 
\beq
C_1 = {\dot M \over 2\pi}, \quad C_2 = {{\cal F}_J \over 2\pi} .
\eeq
The ratio $C_2 / C_1$ has the units of specific angular momentum. In the steady state the disc radial velocity is therefore given by (eq. \ref{radvelaaa})
\beq
U^r = - {W^r_\phi \over U_\phi '} {1 \over \zeta } {\partial \zeta \over \partial r} = - {\dot M W^r_\phi \over U^0 }  \left[  \dot M \int^r {U_\phi'  \over U^0 } \, {\rm d}r + {\cal F}_{\cal J} \right]^{-1} .
\eeq
In the vanishing ISCO stress assumption $W^r_\phi(r_I) = 0$, we require $\zeta(r_I) = 0$. If applied, this condition constrains the angular momentum flux integration constant to be 
\beq
{\cal F}_{{\cal J}, {\cal V}} = - \dot M \int^{r_I} {U_\phi' \over U^0 }\, {\rm d} r .
\eeq
In general this constant can be trivially written as 
\beq
{\cal F}_{\cal J} = {\cal F}_{{\cal J}, {\cal V}} + \delta {\cal F}_{\cal J}
\eeq 
where $\delta {\cal F}_{\cal J} $ can be interpreted as the deviation in the discs total angular momentum flux from that of the vanishing stress solution. The ratio $\delta {\cal F}_{\cal J}/ \dot M$ has the dimensions of a specific angular momentum, and is a number routinely extracted from GRMHD simulations. Most simulations find $\delta {\cal F}_{\cal J}/\dot M \sim 0.01{\cal J}_I-0.1 {\cal J}_I$, where ${\cal J}_I$ is the specific angular momentum of an ISCO circular orbit (e.g., Schaffe et al. 2008, Noble et al. 2010). When $\delta {\cal F}_{\cal J}$ is non-zero we necessarily have a non-zero ISCO stress and  therefore, at the ISCO, we generally have 
\beq\label{urIgenaaa}
U^r (r \rightarrow r_I) = - {\dot M W_I \over U^0 \delta {\cal F}_{\cal J}} \equiv -  u_I ,
\eeq 
where $W_I \equiv W^r_\phi(r_I)$, the ISCO stress. Note that $W_I$ is a function of $\delta {\cal F}_{\cal J}$, and they are not independent parameters. To understand the trans-ISCO velocities dependence on the accretion parameters we must  relate $W_I$ and $\delta {\cal F}_{\cal J}$.  This is the procedure performed in the following two sections. 

\subsection{Explicit profiles of the extra-ISCO disc quantities }
The temperature of the main body of the accretion disc is found by solving the integral 
\begin{multline}
\zeta(r) = {\dot M \over 2\pi} \int^r {U_\phi'  \over U^0 } \, {\rm d}r + {{\cal F}_{\cal J} \over 2\pi}  \\ = {\dot M  c \over 4\pi} \int^r \sqrt{r_g \over r} {r^2 - 6r_gr - 3a^2 + 8a\sqrt{r_g r} \over r^2 - 3r_gr + 2a\sqrt{r_gr}}\, {\rm d}r + {{\cal F}_{\cal J} \over 2\pi}
\end{multline}
with $x \equiv \sqrt{r/r_g}$ and $a_\star = a/r_g$, we have
\beq
\zeta(r) = {\dot M r_g c \over 2 \pi } \int^r  {x^4 - 6x^2 - 3a_\star^2 + 8a_\star x \over x^4 - 3x^2 + 2a_\star x }\, {\rm d}x + {{\cal F}_{\cal J} \over 2\pi} ,
\eeq
which can be solved with standard partial fraction techniques, as was first shown in Page \& Thorne (1974). The roots of the lower cubic 
\beq
x^3 - 3x + 2a_\star = 0,
\eeq
denoted $\{x_\lambda\}$ are 
\beq
x_\lambda = 2 \cos\left[ {1\over 3} \cos^{-1}(-a_\star) - {2\pi\lambda\over3}\right] 
\eeq
with $\lambda = 0, 1, 2$. The solution of the integral is then\footnote{The reader familiar with the classical Page \& Thorne (1974) solution may notice here that our definition of the constants $k_\lambda$ are expressed differently to their Page \& Thorne counterparts. The two sets of constants are in fact entirely equivalent. We write the $k_\lambda$ constants in this manner so that they only depend on $x_\lambda$, allowing the solution to be  written in this compact form.  }
\beq\label{zeta_sol}
\zeta(r) = {\dot M r_g c \over 2 \pi } \left[ x - {3a_\star \over 2} \ln(x) + \sum_{\lambda = 0}^{2} k_\lambda \ln\left|x - x_\lambda\right|\right] + {{\cal F}_{\cal J} \over 2\pi}, 
\eeq
where 
\beq
 k_\lambda \equiv {2 x_\lambda - a_\star(1 + x_\lambda^2)  \over 2(1 - x_\lambda^2)} .
 \eeq
We now choose the physical values of our integration constants. The mass flux $C_1 = {\dot M / 2\pi}$ simply scales the amplitude of each parameter, while $C_2 = {\cal F}_J / 2\pi $ determines the ISCO boundary condition. The vanishing ISCO stress condition enforces 
\begin{multline}
- {{\cal F}_{{\cal J}, {\cal V}} \over \dot M  } \equiv {\cal J}_{\cal V}= \sqrt{GMr_I} \Bigg[ 1- {3a_\star \over 2 x_I} \ln(x_I) \\ + {1\over x_I} \sum_{\lambda = 0}^{2} k_\lambda \ln\left|x_I - x_\lambda\right|\Bigg] ,
\end{multline} 
and the general boundary condition we shall write as 
\begin{multline}
- {{\cal F}_{\cal J} \over \dot M  } \equiv {\cal J}_{\cal V}(1 - \delta_{\cal J})= \sqrt{GMr_I} \Bigg[ 1- {3a_\star \over 2 x_I} \ln(x_I) \\ + {1\over x_I} \sum_{\lambda = 0}^{2} k_\lambda \ln\left|x_I - x_\lambda\right|\Bigg] (1 - \delta_{\cal J} ) .
\end{multline}
This defines $\delta_{\cal J}$, which is our free inner disc stress parameter. We define our inner disc parameter in this manner as $\delta_{\cal J}$ is a number that can be readily determined from GRMHD simulations of accretion discs. Physically $\delta_{\cal J}$ can be thought of as the fraction of the specific angular momentum `passed back' into the disc from the $r < r_I$ region by whatever process is generating the stress at the ISCO. 

Folding this result through to the key radiative temperature profile, we find (eq. \ref{energybalaaa})
\begin{multline}\label{TRaaa}
\sigma T_R^4 = {3 GM\dot M \over 8 \pi r^3} \Bigg[ 1- {3a_\star \over 2 x} \ln(x) + {1\over x} \sum_{\lambda = 0}^{2} k_\lambda \ln\left|x - x_\lambda\right| \\ - {{\cal J}_{\cal V} (1 - \delta_{\cal J})  \over \sqrt{GMr} }  \Bigg] \left[ 1 -{ 3 \over  x^{2} } + { 2 a_\star \over x^{3}} \right]^{-1} .
\end{multline}
To relate the turbulent stress at the ISCO $W_I$ to the radiative temperature at the ISCO, or equivalently the $\delta_{\cal J}$ parameter, one must construct a closure relationship which allows the turbulent stress to be determined from the disc's thermodynamic quantities. The closure relationship most commonly used in the literature is the so-called Shakura-Sunyaev $\alpha$-parameterisation, which we discuss below.

\subsection{ Turbulent stress $\alpha$-parameterisations }
We use the now standard Shakura \& Sunyaev (1973) $\alpha$-parameterisation of the turbulent stress
\beq
W^r_\phi \equiv \alpha r P_g / \rho 
\eeq
where $\alpha$ is a phenomenological parameter of the model assumed to be constant, $P_g$ is the gas pressure, and $\rho$ the disc density. The gas pressure satisfies 
\beq
{P_g \over \rho} = {k_B T_c \over \mu m_p} ,
\eeq
where $T_c$ is the central temperature. We have used the gas pressure $P_g$ to compute the turbulent stress, which prevents the onset of Lightmann-Eardley (1974) instabilities. The reason for adopting this assumption is simple: we wish to construct a solution which is self consistent. If we were instead to set the turbulent stress proportional to the radiation pressure then we would find that any perturbation away from our solution would lead to runaway heating and cooling of the disc. As the $\alpha$-model itself is an ad-hoc simplification of the underlying turbulent evolution, it makes sense to simplify in a self-consistent manner. 

The solutions of our disc equations are extremele optically thick $\kappa\Sigma \gg 1$, and we can therefore relate the central ($T_c$) and radiative ($T_R$, the temperature of the disc at the $\tau = 1$ surface) temperatures of the disc  by 
\beq
T_c^4 = {3 \over 8} \kappa \Sigma T_R^4 ,
\eeq
where $\kappa$ is the disc opacity, which we shall assume to be dominated by electron-scattering opacity $\kappa = \kappa_{\rm es} = {\rm const}$. The radiative temperature is related to the disc quantities in the following way 
\beq
\sigma T_R^4 = {1 \over 2} {\cal F}_{\cal E} =  -{(U^0)^2 \Omega ' \over 2r} \zeta,
\eeq
where $\Omega  \equiv U^\phi / U^0$ is the discs angular velocity. Note that $\Omega' < 0$ at all radii. Remembering that the quantity $\zeta = r \Sigma W^r_\phi / U^0$ is known, these equations suffice to determine the parameter dependence of $W^r_\phi$: 
\beq\label{wrp5}
(W^r_\phi)^5 = -{3 \kappa_{\rm es} k_B^4 \alpha^4 \over 16 \sigma \mu^4 m_p^4} r^2 (U^0)^3 \zeta^2 \Omega ' .
\eeq
\subsection{The full solution and ISCO parameter dependences }
We now have sufficient information to construct the full extra-ISCO accretion solution. We begin with the properties of the crucial trans-ISCO velocity.  

Equation \ref{wrp5} demonstrates that the ISCO values of $W^r_\phi$ and $\zeta$ are related via  $W_I \propto \left(\delta {\cal F}_{\cal J}\right)^{2/5}$, and therefore (from equation \ref{urIgenaaa})
\beq
u_I  \propto W_I^{-3/2} ,
\eeq
and the vanishing ISCO stress limit $W_I = 0$ results in $u_I \rightarrow \infty$. This is unphysical. A finite ISCO stress disc $W_I \neq 0, \delta {\cal F}_{\cal J} \neq 0$, always results in a finite and non-zero trans-ISCO velocity. 

Not only is the trans-ISCO velocity finite and non-zero, but so are all of the discs thermodynamic quantities. To see this, note that upon specifying values for $M$, $a$, $\dot M$ and $\delta_{\cal J}$, eq. \ref{zeta_sol} determines the function $\zeta$, which is non-zero everywhere. Upon specifying a value of the $\alpha$ parameter, the turbulent stress $W^r_\phi$ is then given by expression \ref{wrp5}. Together, these solutions suffice to describe the surface density of the disc $\Sigma = \zeta U^0 / r W^r_\phi$, and once combined with the radiative temperature $T_R$ above (eq. \ref{TRaaa}),  the central temperature $T_c$ of the disc can be computed. This central temperature then determines the scale height $H$, the gas $P_g$ and radiation $P_r$ pressures of the disc, and the disc density $\rho$. 

The key point here is that only one disc quantity naturally goes to zero at the ISCO: the angular momentum gradient of a circular orbit $U_\phi'$. No other disc quantity depends explicitly on $U_\phi'$, merely on its integral through their dependence on $\zeta$. Thus, a disc with a finite trans-ISCO velocity must have non-zero values for all of its thermodynamic quantities at the ISCO. 

\section{The ballistic flow solutions  of the intra-ISCO region }\label{ballistic_appendix}
For this Appendix we use geometric units $G = c = 1$, so as to de-clutter the lengthy governing equations. In this choice of units both $M$ and $a$ have dimensions of length. 

For circular orbits at radii $r \geq r_I$ the covariant 4-momenta $U_\phi$ and $U_0$ are calculated in the usual manner, and are proportional to the conserved angular momentum and energy, respectively:
\begin{align}
U_\phi &= {(Mr)^{1/2}\over {\cal D}}(1+a^2/r^2-2aM^{1/2}/r^{3/2}),  \\
 U_0 &= - {1\over {\cal D}} (1-2M/r +aM^{1/2}/r^{3/2}) , \\
  {\cal D}^2 &= 1 -3M/r +2aM^{1/2}/r^{3/2}.
  \end{align}
To ensure the stability of circular orbits we require $\partial_r U_\phi > 0$, the relativistic analogue of the Rayleigh criterion.   Evaluating this gradient, we find 
\begin{equation}
{\partial U_\phi \over \partial r}  =
{ M^{1/2}\over 2r^{1/2}{\cal D}^3}\left(1 +{aM^{1/2}\over r^{3/2}}\right)  \left( 1 - {6M\over r} -{3a^2\over r^2} +{8aM^{1/2}\over r^{3/2}}\right) .   
\end{equation}
The factor ${\cal D}^3$  is positive everywhere over the spin domain $-1<a/M<1$ of interest, so the expression passes through zero at $r=r_I$,  where the final factor vanishes, i.e.:
\begin{equation}\label{a}
r_I^2  =    6Mr_I-8a\sqrt{Mr_I} +3a^2  = {2Mr_I\over 3}+{16Mr_I\over 3}\left( 1 - {3a\over 4\sqrt{Mr_I}}\right)^2
\end{equation}
The second equality will be especially convenient in what follows below. We define the constants of motion
\beq
J = U_\phi(r_I), \quad \gamma = - U_0(r_I),
\eeq
with a notation that reminds us of their association with the conserved angular momentum and energy.    If we now use equation (\ref{a}) to substitute for $r_I^2$ in $J$ and $\gamma$, we find that the resulting numerators and denominators in these expressions factor cleanly, leading to a further simplification:
\begin{align}
J &= 2\sqrt{3}M\left( 1 - {2a\over 3\sqrt{Mr_I} }\right) , \label{jjj} \\ \gamma &= {4\over3}\sqrt{3}\left(M\over r_I\right)^{1/2} \left( 1 -{3a\over 4\sqrt{Mr_I}}\right)  = \left(1- {2M\over 3r_I}\right)^{1/2}, \label{gam}
\end{align}
where in the final $\gamma$ equality we have made use of equation (\ref{a}).

General relativistic dynamics permits radial motion in orbits whose angular momentum and energy correspond to an exactly circular orbit.  
The governing equation is easily stated in formal terms:
\beq
g_{rr} (U^r)^2 +U^0U_0 +U^\phi U_\phi =-1, 
\eeq
but its direct solution is a matter of no little algebraic complexity.   
Expanding, expressing all non-radial 4-velocities in terms of $J$ and $\gamma$, and multiplying through by $1/g_{rr}$, we have
\begin{multline}\label{full}
 (U^r)^2 + {J\over r^2} \left( {2Ma\gamma \over r} + \left(1 - {2M \over r}\right) J \right) \\ - {\gamma\over r^2} \left[\left(r^2 + a^2 + {2Ma^2 \over r}\right) \gamma - {2MaJ \over r}\right] \\ = -1-{a^2\over r^2} +{2M\over r}.
\end{multline}
Before presenting the results of an algebraic simplification of equation \ref{full}, note that the ISCO radius should correspond to a triple root of $U^r$. This can be understood by noting that equation \ref{full} is of the form $(U^r)^2 + V_{\rm eff}(r) = 0$, which defines an `effective' potential $V_{\rm eff}$ which is cubic in $1/r$.  This may be factored to take the form
\beq
 V_{\rm eff}(r) = - V_0 \left({r_1\over r} - 1\right)\left({r_2\over r}- 1\right)\left({r_3\over r}-1\right),
\eeq 
where $r_1, r_2$ and $r_3$ are the general (possibly complex) roots of $V_{\rm eff}$. 
For a circular orbit of radius $r=r_c$,  both $V_{\rm eff}(r_c) = 0$ and $\partial_r V_{\rm eff}(r_c) = 0$, and there will be a double root of the polynomial.    For the particular case of the last stable circular orbit, there is an additional condition, $\partial^2_r V_{\rm eff}(r_I) = 0$.  Thus, we expect $r_I$ to be a {\it triple} root of $U^r$.   The normalisation constant $V_0$ may be found by going back to equation (\ref{full}) and taking the formal limit $r\rightarrow \infty$. We find
\beq
V_0= - (U^r)^2 \rightarrow 1- \gamma^2 =   {2M \over 3 r_I} ,
\eeq   
which leads directly to our final equation for (inward moving) $U^r$:
\beq\label{flow}
U^r \equiv  {{\rm d}r\over {\rm d}\tau} = - \sqrt{2M\over 3r_I} \left({r_I \over r} - 1\right)^{3/2}.
\eeq
This result can be confirmed by the much more involved direct substitution of equations \ref{jjj} \& \ref{gam} into \ref{full}, making repeat use of equation \ref{a} to remove powers of $a^2$ and higher. 

\section{The vanishing of the magnetic contribution to the entropy equation}\label{mag_proof} 
We wish to prove that 
\begin{multline}
U_\nu \nabla_\mu T^{\mu\nu}_{\rm mag} =  {U_\nu \over 4\pi\mu_0} \Bigg({b^2 \over c^2} U^\mu \nabla_\mu U^\nu + {b^2 \over c^2} U^\nu \nabla_\mu U^\mu  + \\  \left({U^\mu U^\nu \over c^2} + {1\over 2} g^{\mu\nu}\right) \nabla_\mu b^2 - b^\mu\nabla_\mu b^\nu - b^\nu \nabla_\mu b^\mu  \Bigg] = 0 .
\end{multline}
Some of these terms vanish trivially, a result of the identities 
\beq
U_\nu b^\nu = 0, \quad U_\nu U^\nu = - c^2, \quad U_\nu \nabla_\mu U^\nu = 0. 
\eeq
The non-trivial terms which remain are 
\beq
4\pi\mu_0 U_\nu \nabla_\mu T^{\mu\nu}_{\rm mag} = - b^2 \nabla_\mu U^\mu - {1\over 2}U^\mu \nabla_\mu b^2 - b^\mu U_\nu \nabla_\mu b^\nu .
\eeq
The induction equation in full general relativity is (Misner, Thorne and Wheeler 1973)
\beq
\nabla_\mu \left(U^\mu b^\nu - b^\mu U^\nu \right) = 0 ,
\eeq
which combines the classical induction equation and the no monopoles condition. This result implies 
\beq
b^\nu \nabla_\mu U^\mu = U^\nu \nabla_\mu b^\mu + b^\mu \nabla_\mu U^\nu - U^\mu \nabla_\mu b^\nu ,
\eeq
and therefore 
\beq
b^2 \nabla_\mu U^\mu = b^\mu b_\nu \nabla_\mu U^\nu - b_\nu U^\mu \nabla_\mu b^\nu . 
\eeq
Substituting this result in above leaves 
\begin{multline}
4\pi \mu_0U_\nu \nabla_\mu T^{\mu\nu}_{\rm mag} = -b^\mu b_\nu \nabla_\mu U^\nu + b_\nu U^\mu \nabla_\mu b^\nu   \\ - {1\over 2}U^\mu \nabla_\mu b^2 - b^\mu U_\nu \nabla_\mu b^\nu . 
\end{multline}
Using 
\beq
{1\over 2} \nabla_\mu b^2 = b_\nu \nabla_\mu b^\nu ,
\eeq
leads to a simple cancellation  
\begin{equation}
4\pi \mu_0U_\nu \nabla_\mu T^{\mu\nu}_{\rm mag} = -b^\mu b_\nu \nabla_\mu U^\nu   - b^\mu U_\nu \nabla_\mu b^\nu. 
\end{equation}
Trivial simplifications then imply 
\begin{multline}
4\pi \mu_0U_\nu \nabla_\mu T^{\mu\nu}_{\rm mag} = - b^\mu \left( b_\nu \nabla_\mu U^\nu + U_\nu \nabla_\mu b^\nu \right) \\ = -b^\mu \nabla_\mu (b_\nu U^\nu) = 0 ,
\end{multline}
which completes the proof. 

\section{The formal non-intersection of the intra-ISCO inspirals }
It is interesting to note that the 2-dimensional ($r$ and $\phi$) intra-ISCO inspirals do not intercept, and there will therefore not be collisions of fluid elements within the ISCO. The formal radial  velocity of the intra-ISCO inspiral is 
\beq
U^r = - c \sqrt{2r_g \over 3 r_I} \left( {r_I \over r} - 1\right)^{3/2} , 
\eeq
while the azimuthal velocity is 
\beq
U^\phi = - g^{\phi 0} \gamma + g^{\phi\phi} J = {2r_g\gamma a c + J(r -2r_g) \over r(r^2 - 2r_gr + a^2)} .
\eeq
The key points to note here are that,  (1) both the radial and azimuthal velocities depend only on radius, and that (2) the radial velocity increases inwards. Consider two fluid elements located at $r_1$ and $r_2$,  separated by $\delta r \equiv r_1 - r_2 > 0$. Then 
\beq
|U^r(r_2)| > |U^r(r_1)| \to {{\rm d} \over {\rm d}\tau } \delta r =  |U^r(r_2)| - |U^r(r_1)| > 0, 
\eeq
and the fluid elements move {\it further apart}. Similarly, two fluid elements separated by $\delta \phi \equiv \phi_2 - \phi_1$, but at the same radius,  have 
\beq
U^\phi(\phi_1) = U^\phi(\phi_2) \to {{\rm d} \over {\rm d}\tau } \delta \phi =  U^\phi(\phi_2) - U^\phi(\phi_1) = 0, 
\eeq
and the fluid elements maintain their angular separation. Any initial offset $(\delta r, \delta \phi)$ between fluid elements therefore strictly {\it grows}, and the inspirals do not formally intercept. 

The arguments presented here will break down if turbulence produces  velocities of the form 
\beq
U^r = - c \sqrt{2r_g \over 3 r_I} \left( {r_I \over r} - 1\right)^{3/2}  + \delta U^r(r, \phi), 
\eeq
and 
\beq
U^\phi = {2r_g\gamma a c + J(r -2r_g) \over r(r^2 - 2r_gr + a^2)} + \delta U^\phi(r, \phi) ,
\eeq
which is entirely possible. Then, in the small radial region near to the ISCO where $U^r \ll c$, the random fluctuations in the 4-velocity of the fluid elements could produce some fluid elements which are moving faster than others upstream, leading to  collisions, dissipation, heating, and light. However, once the fluid elements move significantly from the ISCO, their velocities will be completely dominated by their geodesic flow, and the solutions will again cease to intercept. 

\section{Parameter dependence of the intra-ISCO thermodynamics}\label{parameter_plots_C}
In this appendix we present the effect of varying the black hole and disc parameters on the intra-ISCO region. In these Figures we fix the remaining parameters to that of Fig. \ref{fig4}, which we remind the reader was constructed with $M = 10 M_\odot$, $\alpha = 0.1$, $a = 0$, $\delta_{\cal J} = 10^{-4}$, and $\dot M = 0.1 \dot M_{\rm edd}$. 

\subsection{Varying the black hole mass}
\begin{figure*}
    \includegraphics[width=.95\linewidth, height=1.25\linewidth]{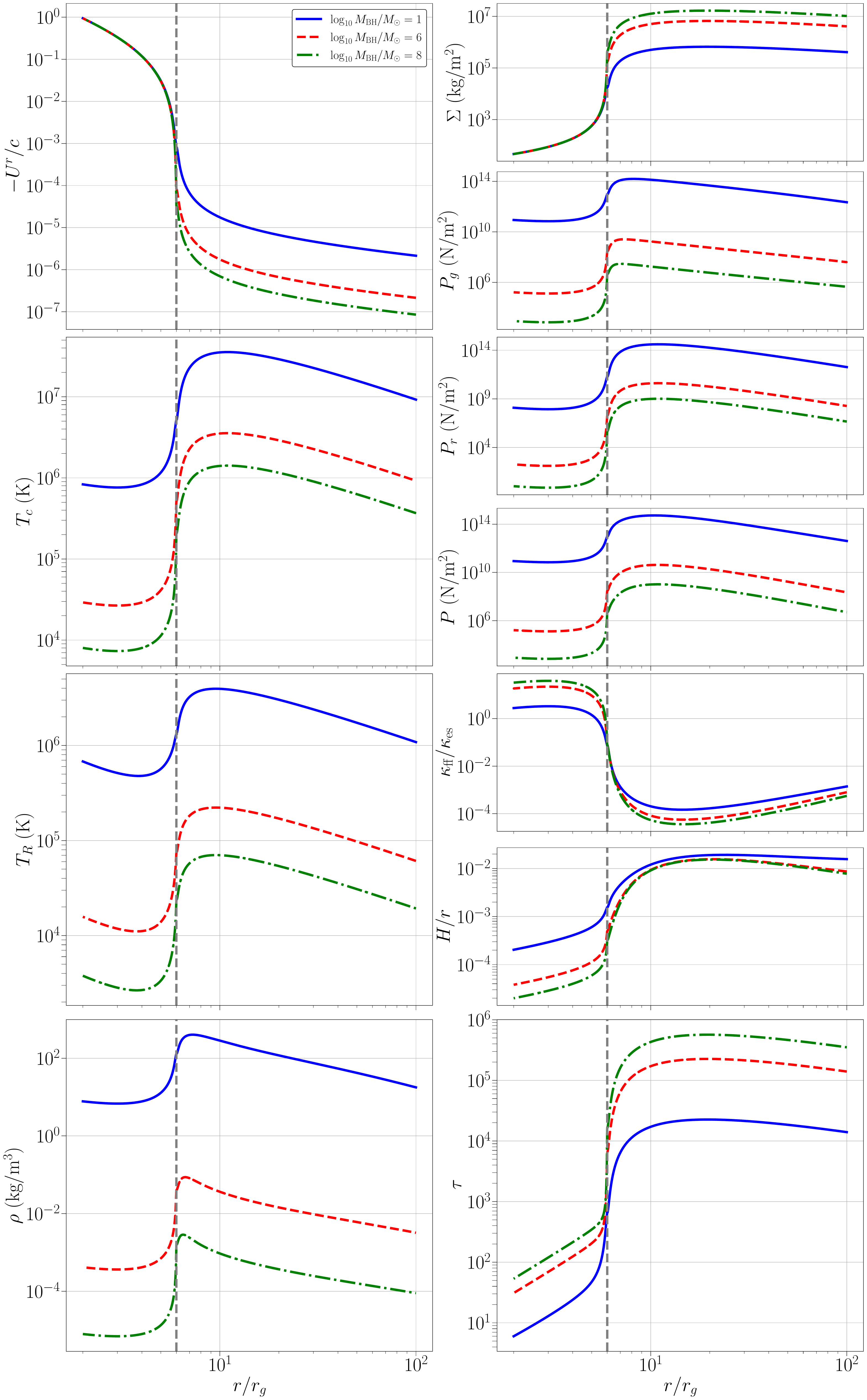} 
     \caption{ As in Fig. \ref{fig4}, except for a black hole masses of $M = 10, 10^6$ and $10^8 M_\odot$ (denoted in Figure legend). We note that, in addition to the different intrinsic parameter scales caused by different black hole masses,  the radiative temperature falls more rapidly as a fraction of its ISCO value for super massive black holes, owing to the dominance of the free-free opacity in their discs.  } 
 \label{figc1}
\end{figure*}

First we consider the effect of varying the black hole mass on the intra-ISCO flow. The solution displayed in Fig. \ref{fig4} was for a black hole mass typical of those in a Galactic X-ray binary.  We can also trivially compute the global accretion solutions for supermassive black hole discs, which we display in Fig. \ref{figc1} for the case of $M = 10, 10^6$ and $10^8 M_\odot$ (all other free-parameters are kept the same as Fig. \ref{fig4}). Qualitatively the solutions for supermassive black holes are identical to those of solar mass black holes, but we do note that the exact solutions differ in the quantitative details. In particular the radiative temperature falls more rapidly across the intra-ISCO region of super massive black holes, owing to the dominance of the free-free opacity in these discs. As was discussed in the proceeding section, the intra-ISCO solutions display a ``hotter discs stay hotter'' behaviour which is relevant in understanding the differing properties of stellar-mass and supermassive black hole discs.

\subsection{Varying the black hole spin}
\begin{figure*}
    \includegraphics[width=.95\linewidth, height=1.25\linewidth]{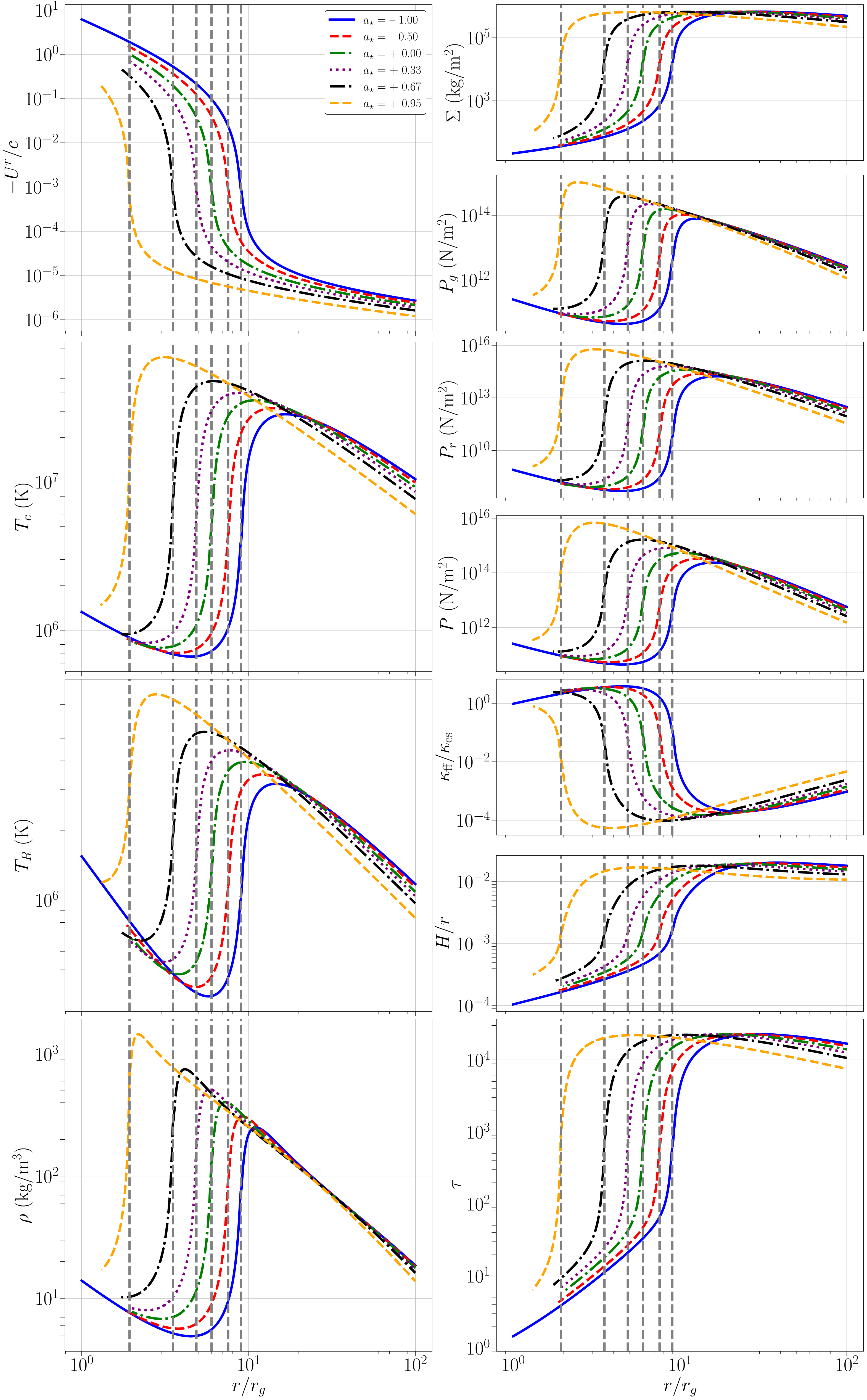} 
     \caption{ As in Fig. \ref{fig4}, except for a variety of black hole spins $a_\star \equiv a/r_g = -1, -0.5, 0, 0.33, 0.67, 0.95$ (denoted in Figure legend). The ISCO radius of each disc is denoted by a dashed grey line, with higher (positive) spins having ISCO radii closer to $r = 0$. The intra-ISCO region is stopped at the event horizon of each black hole. We do not plot the change in parameter dependence in the free-streaming region in this Figure (relevant for $a_\star = -0.5, -1$).    } 
 \label{figc2}
\end{figure*}

In Fig. \ref{figc2} we construct the global accretion solution for a variety of black hole spins $a_\star \equiv a/r_g = -1, -0.5, 0, 0.33, 0.67, 0.95$. Here all other parameters are kept the same as Fig. \ref{fig4}. The ISCO radius of each disc is denoted by a dashed grey line, with higher (positive) spins having ISCO radii closer to $r = 0$. The intra-ISCO region is stopped at the event horizon of each black hole. We do not plot the change in parameter dependence in the free-streaming region in this Figure (relevant for $a_\star = -0.5, -1$), see section 7 for a discussion of the solutions in that regime.  

Even though the intra-ISCO solutions only depend on black hole spin implicitly through the ISCO radius, the black hole spin does qualitatively effect some of the properties of these discs. This is a result of the event horizon taking on progressively smaller fractions of the ISCO radius as the spin is made increasingly negative. Note that there is a region characterised by $|U^r| > c$ outside of the black hole event horizon for sufficiently large retrograde spins. This result is not in contradiction with causality. The radial velocity is defined as $U^r = {\rm d}r/{\rm d}\tau$, where $r$ is a metric coordinate, not a physical distance. 

\subsection{Varying the accretion parameters}
\begin{figure*}
    \includegraphics[width=.95\linewidth, height=1.25\linewidth]{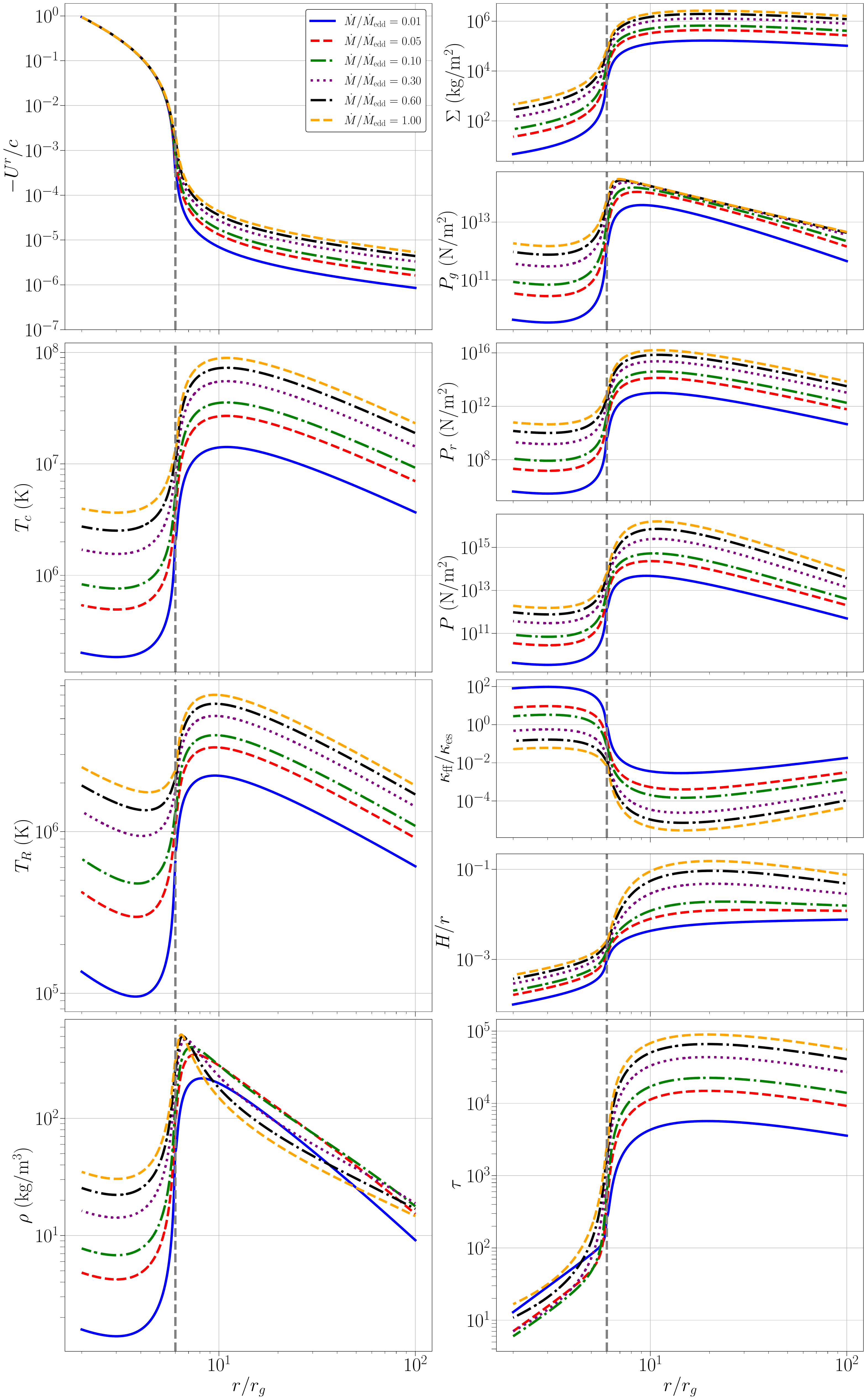} 
     \caption{ As in Fig. \ref{fig4}, except for accretion rates of $\dot M/\dot M_{\rm edd} = 0.01, 0.05, 0.1, 0.3, 0.6, 1$ (denoted in Figure legend). These solutions highlight the ``hotter discs stay hotter'' character of the intra-ISCO region. Note that the thin-disc approximations fail for the highest accretion rates, and these solutions should be considered purely formal.      } 
 \label{figc3}
\end{figure*}

In Fig. \ref{figc3} we construct the global accretion solution for a variety of mass accretion rates  $\dot M/\dot M_{\rm edd} = 0.01, 0.05, 0.1, 0.3, 0.6, 1$. Here all other parameters are kept the same as Fig. \ref{fig4}.  These solutions highlight the ``hotter discs stay hotter'' character of the intra-ISCO region. Note that the thin-disc approximations fail for the highest accretion rates, and these solutions should be considered purely formal.  We note that increasing the $\alpha$-parameter of the disc has a similar qualitative effect as increasing the accretion rate. 

\subsection{Varying the ISCO stress}
\begin{figure*}
    \includegraphics[width=.95\linewidth, height=1.25\linewidth]{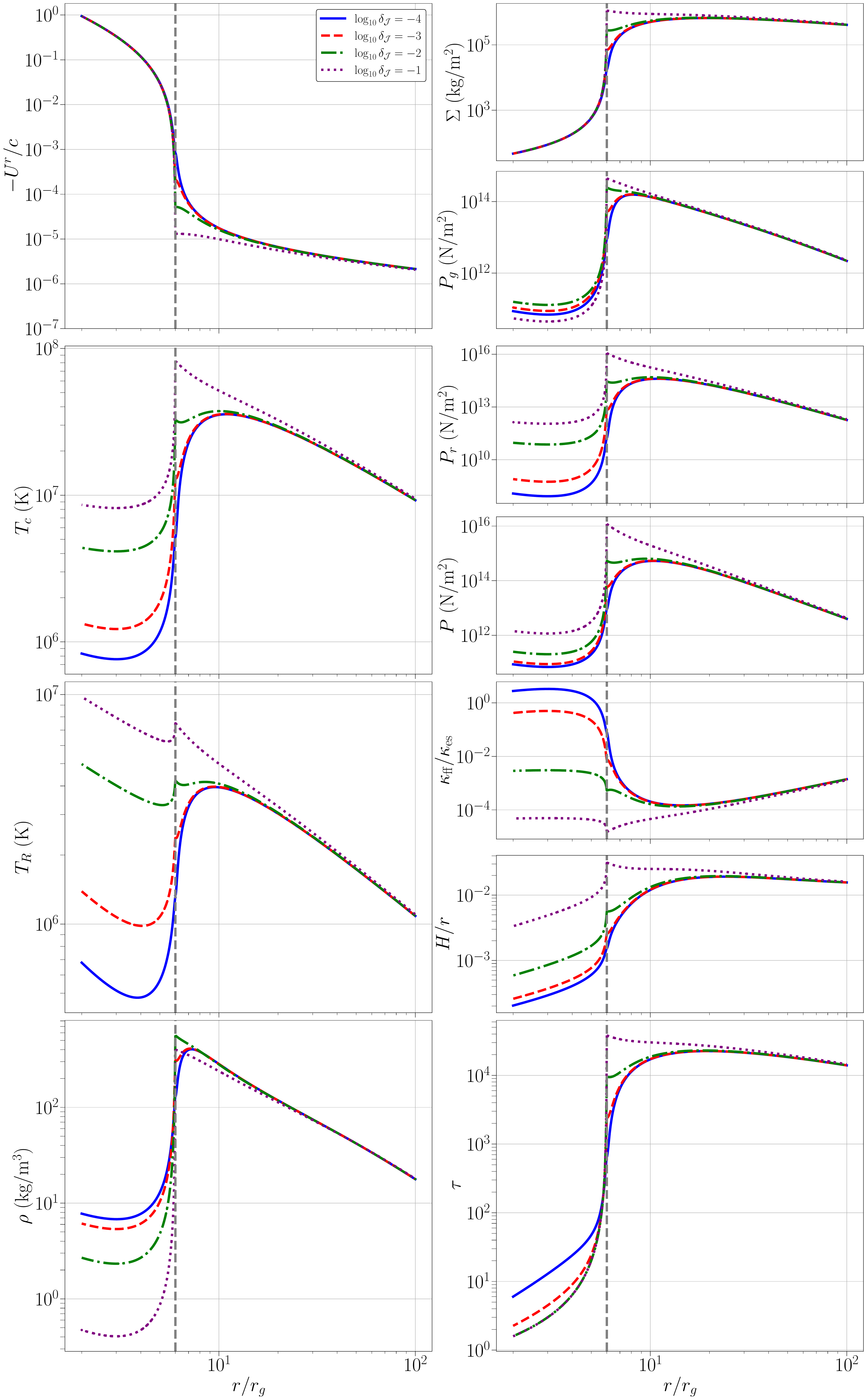} 
     \caption{ As in Fig. \ref{fig4}, except for ISCO stress parameters $\delta_{\cal J} = 10^{-4}, 10^{-3}, 10^{-2}, 10^{-1}$ (denoted in Figure legend). The value of the ISCO stress is clearly a parameter of upmost importance, and dictates much of the properties of the intra-ISCO region.  In particular, the radiative temperature of the disc may peak {\it within} the ISCO if the ISCO stress is sufficiently large.   } 
 \label{figc4}
\end{figure*}

In Fig. \ref{figc4} we construct the global accretion solution for a variety of ISCO stress parameters $\delta_{\cal J} = 10^{-4}, 10^{-3}, 10^{-2}, 10^{-1}$. Here all other parameters are kept the same as Fig. \ref{fig4}.  The value of the ISCO stress is clearly a parameter of upmost importance, and dictates much of the properties of the intra-ISCO region.  In particular, the radiative temperature of the disc may peak {\it within} the ISCO if the ISCO stress is sufficiently large.  We note that $\delta_{\cal J} = 10^{-1}$ is the value for the ISCO stress found in the simulations of Noble et al. (2010), and  may therefore represent the most realistic accretion solution. 

\label{lastpage}

\end{document}